\documentclass[10pt,letterpaper]{article}
\usepackage[top=0.85in,left=1.5in,footskip=0.75in,marginparwidth=2.5in]{geometry}

\usepackage[utf8]{inputenc}

 \usepackage{natbib}
\usepackage{changepage}

\usepackage[aboveskip=1pt,labelfont=bf,labelsep=period,singlelinecheck=off]{caption}

\makeatletter
\renewcommand{\@biblabel}[1]{\quad#1.}
\makeatother
\usepackage[section]{placeins}
\usepackage{lastpage,fancyhdr,graphicx}
\usepackage{epstopdf}
\fancyheadoffset[L]{1.25in}
\fancyfootoffset[L]{1.25in}

\usepackage{color}

\definecolor{Gray}{gray}{.25}

\usepackage{graphicx}

\usepackage{sidecap}

\usepackage{wrapfig}
\usepackage[pscoord]{eso-pic}
\usepackage[fulladjust]{marginnote}
\reversemarginpar

\usepackage[T1]{fontenc} % Output font encoding for international characters
\usepackage{palatino} % Use the Palatino font by default
\usepackage[flushleft]{threeparttable}
\usepackage{amsthm}

\usepackage{algorithm} % psedoudocode
\usepackage{algorithmic} % psedoudocode
\usepackage{placeins}  % avoid floating
\usepackage[autostyle=true]{csquotes} % Required to generate language-dependent quotes in the bibliography
\usepackage{graphicx}
\usepackage{amsmath}
\usepackage{amssymb}
\usepackage{tikz}
\usetikzlibrary{bayesnet}
\usetikzlibrary{fit}
\usepackage{stackengine}
\def\shft#1{\stackunder[15pt]{}{\kern50pt #1}}

\usepackage{parskip}
\begin{document}
\vspace*{0.35in}

% title goes here:
\begin{flushleft}
{\Large
\textbf\newline{Naive Dictionary On Musical Corpora: \\ From Knowledge Representation To Pattern Recognition}
}
\newline
% authors go here:
\\
Qiuyi Wu\textsuperscript{1,*},
Ernest Fokou\'e\textsuperscript{1}
\\
\bigskip
\bf{1} School of Mathematical Science, Rochester Institute of Technology, Rochester, New York, USA
\\

\bigskip
* wu.qiuyi@mail.rit.edu

\end{flushleft}

\section*{Abstract}
In this paper, we propose and develop the novel idea of treating musical sheets as literary documents in the traditional text analytics parlance, to fully benefit from the vast amount of research already existing in statistical text mining and topic modelling. We specifically introduce the idea of representing any given piece of music as a collection of  "musical words"  that we codenamed "muselets", which are essentially musical words of various lengths. Given the novelty and therefore the extremely difficulty of properly forming a complete version of a dictionary of muselets, the present paper focuses on a simpler albeit naive version of the ultimate dictionary, which we refer to as a Naive Dictionary because of the fact that all the words are of the same length. We specifically herein construct a naive dictionary featuring a corpus made up of African American, Chinese, Japanese and Arabic music, on which we perform both topic modelling and pattern recognition. Although some of the results based on the Naive Dictionary are reasonably good, we anticipate phenomenal predictive performances once we get around to actually building a full scale complete version of our intended dictionary of muselets.

% the * after section prevents numbering
\section{Introduction}
Music and text are similar in the way that both of them can be regraded as information carrier and emotion deliverer. People get daily information from reading newspaper, magazines, blogs etc., and they can also write diary or personal journal to reflect on daily life, let out pent up emotions, record ideas and experience. Composers express their feelings through music with different combinations of notes,  diverse tempo\footnote[1]{In musical terminology, tempo ("time" in Italian), is the speed of pace of a given piece.},  and dynamics levels\footnote[2]{In music, dynamics means how loud or quiet the music is.}, as another version of language. \\

This paper explores various aspects of statistical machine learning methods for music mining with a concentration on music pieces from Jazz legends like Charlie Parker and Miles Davis. We attempt to create a Naive Dictionary analogy to the language lexicon. That is to say, when people hear a music piece, they are hearing the audio of an essay written with "musical words", or "muselets". The target of this research work is to create homomorphism between musical and literature. Instead of decomposing music sheet into a collection of single notes, we attempt to employ direct seamless adaptation of canonical topic modeling on words in order to "topic model" music fragments. \\

One of the most challenging components is to define the basic unit of the information from which one can formulate a soundtrack as a document. Specifically, if a music soundtrack were to be viewed as a document made up of sentences and phrases, with sentences defined as a collection of words (adjectives, verbs, adverbs and pronouns), several topics would be fascinating to explore: 
\begin{itemize}
\item What would be the grammatical structure in music?
\item What would constitute the jazz lexicon or dictionary from which words are drawn? 
\end{itemize}

All music is story telling as assumption. It is plausible to imagine every piece of music as a collection of words and phrases of variable lengths with adverbs and adjectives and nouns and pronouns. 
\begin{align*}
\varphi: \textit{ musical sheet }\rightarrow \textit{ bag of music words}
\end{align*}
The construction of the mapping $\varphi$ is non-trivial and requires deep understanding of music theory. Here several great musicians offer insights on the complexity of $\varphi$ from their perspectives, to explain about the representation of the input space, namely, creating a mapping from music sheet to collection of music "words" or "phrases":

\begin{itemize}
\item \textit{"These are extremely profound questions that you are asking here. I think I’m interested in trying. But you have opened up a whole lot of bigger questions with this than you could possibly imagine."} (Dr. Jonathan Kruger, personal communication with Dr. Ernest Fokoue, November 24, 2018).
\item \textit{"Your music idea is fabulous but are you sure that nothing exists? Do you know "band in a box? It is a software in which you put a sequence of chords and you get an improvisation 'à la manière de'. You choose amongst many musicians so they probably have the dictionary to play as Miles, Coltrane, Herbie, etc."} (Dr. Evans Gouno, personal communication with Dr. Ernest Fokoue, November 05, 2018).
\item \textit{Rebecca Ann Finnangan Kemp mentioned building blocks of music when it comes to music words idea.} (personal communication with Dr. Ernest Fokoue, November 20, 2018).
\end{itemize}

The concept of \textit{notes} is equivalent to \textit{alphabet}, which can be extended as below:  
\begin{itemize}
\item literature word $\equiv$ mixture of the 26 alphabets
\item music word $\equiv$ mixture of the 12 musical notes 
\end{itemize}

Since notes are fundamental, one can reasonably consider input space directly isomorphic to the 12 notes. \\

%-----------------------------------
%	SECTION 2
%-----------------------------------
\section{Related Work}

\begin{table}[H]
\centering
\caption{Comparison between Text and Music in Topic Modeling}\label{compare}
\begin{threeparttable}
 \begin{tabular}{ |c||c|c|c|c|c| } 
\hline
 Text & letter &word & topic & document  & corpus \\ [0.5ex]
 \hline
 Music & note & notes* & melody & song & album \\ 
 \hline
\end{tabular}
\begin{tablenotes}
 \scriptsize
\item[*] \textbf{a series of notes in one bar can be regarded as a "word"}
\end{tablenotes}
\end{threeparttable}
\end{table}

 \begin{figure}[H]
 \centering
  \includegraphics[scale=0.5]{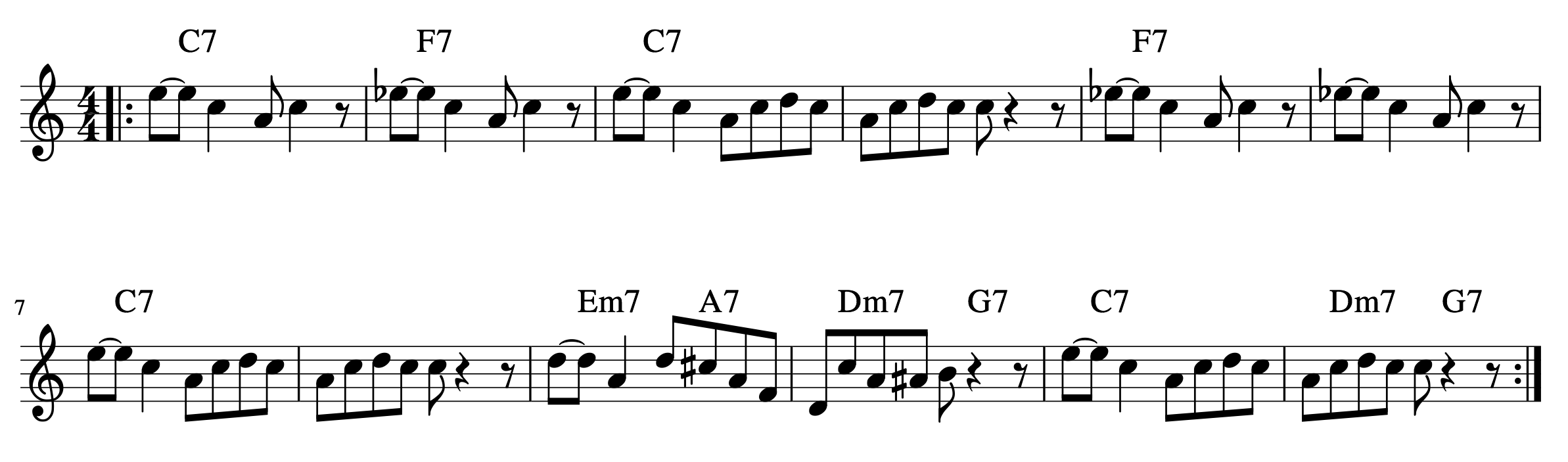}
  \caption{Piece of Music Melody}
 \end{figure}

Compared with the role of text in Topic Modeling as showed in Table \ref{compare}, we treat a series of notes as "word", can also be called as "term", as single note could not hold enough information for us to interpret, specifically, we treat notes in one bar\footnote[3]{In musical notation, a bar (or measure) is a segment of time corresponding to a specific number of beats in which each beat is represented by a particular note value and the boundaries of the bar are indicated by vertical bar lines.} as one "term". Melody\footnote[4]{Harmony is formed by consecutive notes so that the listener automatically perceives those notes as a connected series of notes.} plays the role of "topic", and the melodic materials give the shape and personality of the music piece. "Melody" is also referred as "key-profile" by \citet{Hu09} in their paper, and this concept was based on the key-finding algorithm from \citet{ks1990} and the empirical work from \citet{kk1982}. The whole song is regarded as "document" in text mining, and a collection of songs called album in music could be regarded as "corpus" in text mining. \\

\begin{figure}[H]
   \begin{minipage}{0.5\textwidth}
     \centering
  \includegraphics[width=1\linewidth]{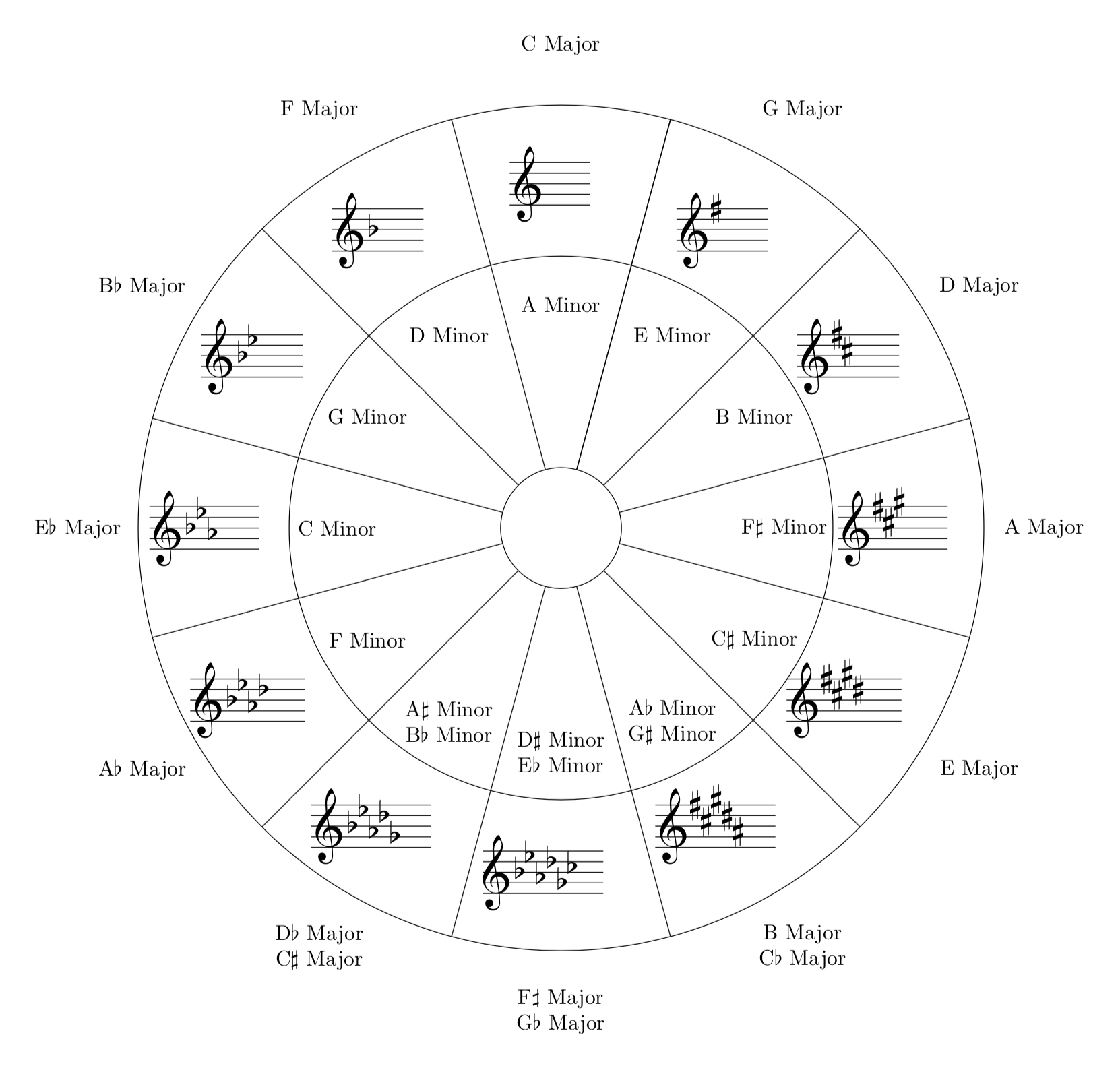}
   \end{minipage}\hfill
   \begin{minipage}{0.5\textwidth}
     \centering
  \includegraphics[width=1\linewidth]{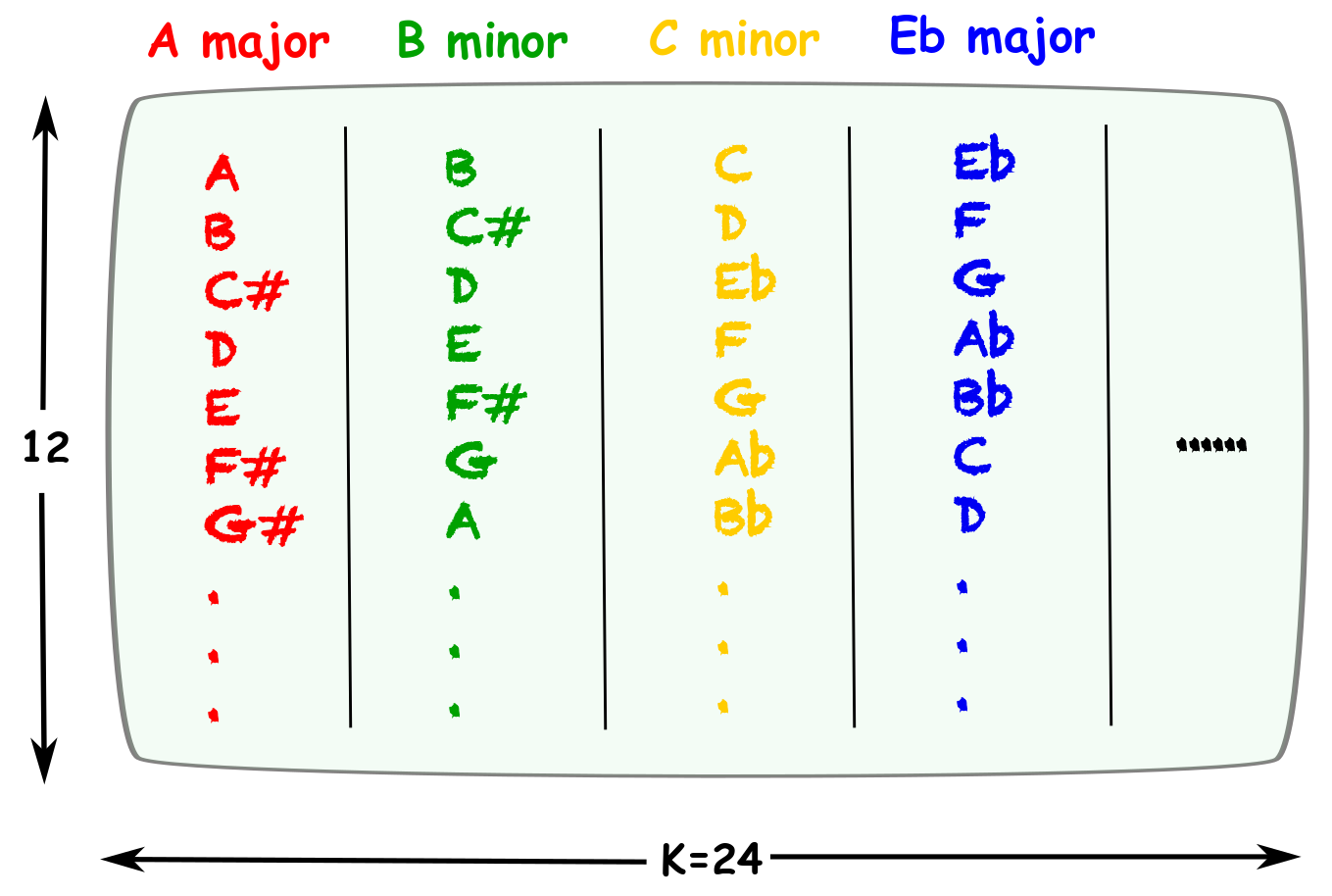}
   \end{minipage}
\caption{Circle of Fifths (left) and Key-profiles (right)}\label{fig:fifths}
\end{figure}

Specifically, "key-profile" is chromatic scale showed geometrically in Figure \ref{fig:fifths} Circle of Fifths plot containing 12 pitch classes in total with major key and minor key respectively, thus there are totally 24 key-profiles, each of which is a 12-dimensional vector. The vector in the earliest model in \citet{long1971} uses indicator with value of 0 and 1 to simply determine the key of a monophonic piece.
E.g. C major key-profile:
$$ [1,0,1,0,1,1,0,1,0,1,0,1]$$
As showed in the figures below, \citet{ks1990} judge the key in a more robust way. Elements in the vector indicate the stability of each pitch-class corresponding to each key. Melody in the same key-profile would have similar set of notes, and each key-profile is a distribution over notes. \\

\begin{figure}[H]
   \begin{minipage}{0.5\textwidth}
     \centering
     \includegraphics[width=1\linewidth]{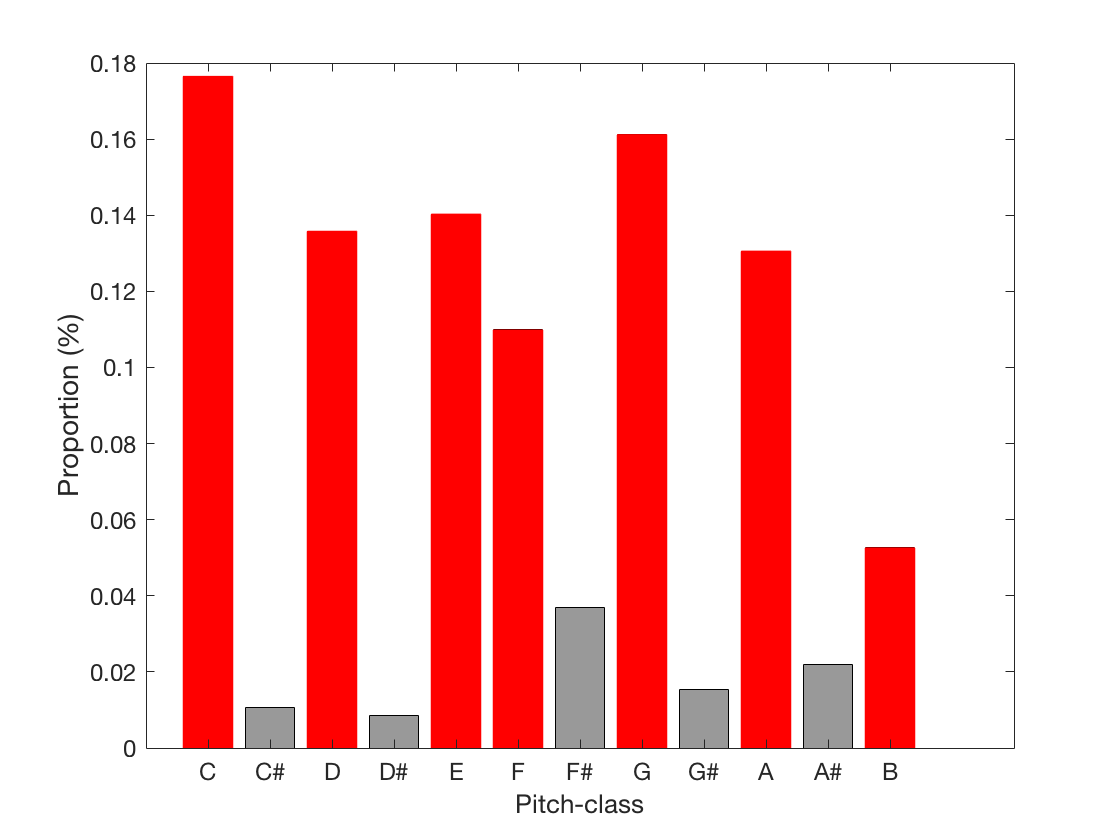}
     \caption{C major key-profile}\label{Fig:Cmajor}
   \end{minipage}\hfill
   \begin{minipage}{0.5\textwidth}
     \centering
     \includegraphics[width=1\linewidth]{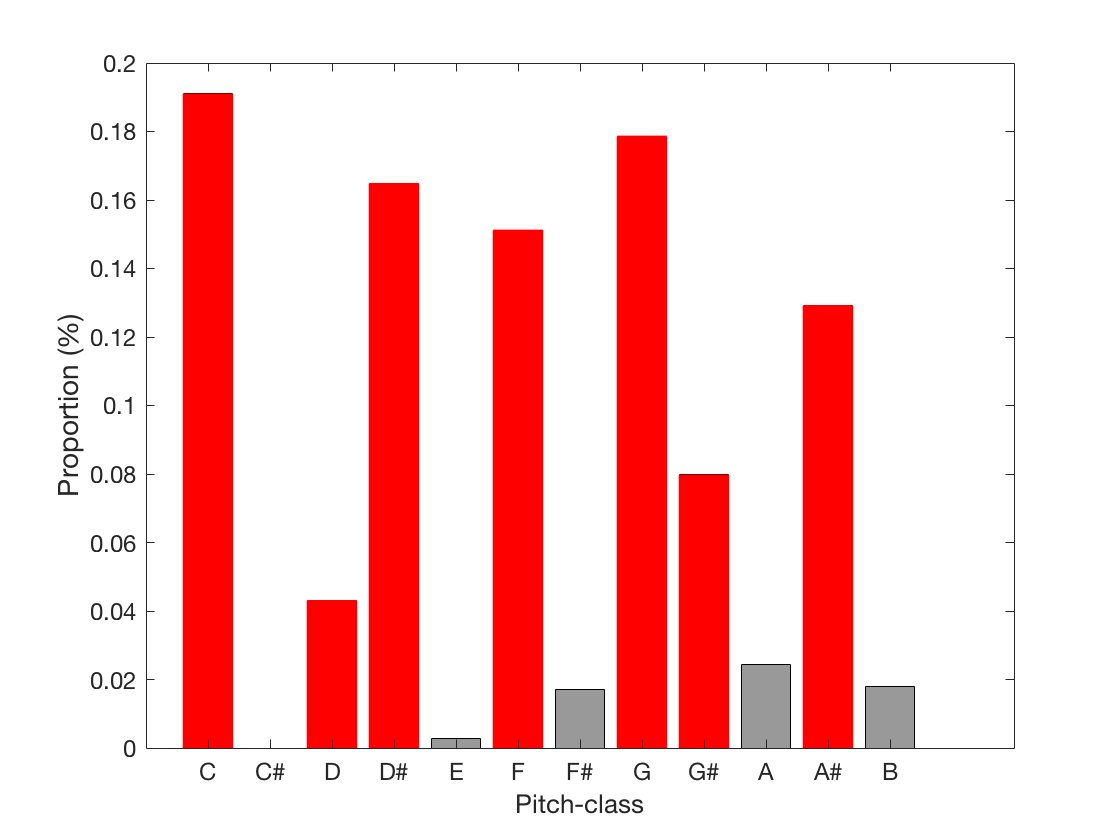}
     \caption{C minor key-profile}\label{Fig:Cminor}
   \end{minipage}
\end{figure}

Figure \ref{Fig:Cmajor} shows the pitch-class distribution of C Major \textit{Piano Sonata No.1, K.279/189d (Mozart, Wolfgang Amadeus)} using K-S key-finding algorithm, and we can see all natural notes: C, D, E, F, G, A, B have high probability to occur than other notes. 
Figure \ref{Fig:Cminor} shows the pitch-class distribution of C Minor \textit{BWV.773 No. 2 in C minor (Bach, Johann Sebastian)} and again we can see specific notes typical for C Minor with higher probability: C, D, D$\sharp$, F, G, G$\sharp$, and A$\sharp$.

Usually different scales could bring different emotions. Generally, major scale arouse buoyant and upbeat feelings while minor scales create dismal and dim environment. Details for emotion and mood effects from musical keys would be presented in later section.

\section{Representation}
We mainly studied symbolic music in \texttt{mxl} format in this research work. The data are collected from MuseScore\footnote[2]{MuseScore: https://musescore.org/en} containing music pieces from different musicians and genres. Specifically, we collect music pieces from 3 different music genres, i.e.: Chinese songs, Japanese songs, Arabic songs. For Jazz music we collect work from 7 different musicians, i.e.: Duke Ellington, Miles Davis, John Coltrane, Charlie Parker, Louis Armstrong, Bill Evans, Thelonious Monk. 

\begin{itemize}
\item Transfer \texttt{mxl} file to \texttt{xml} file 
\item Use \texttt{mxl} files to extract notes in each measure 
\item Create matrices based on the extracted notes
\end{itemize}
\begin{figure}[H]
 \centering
  \includegraphics[scale=0.4]{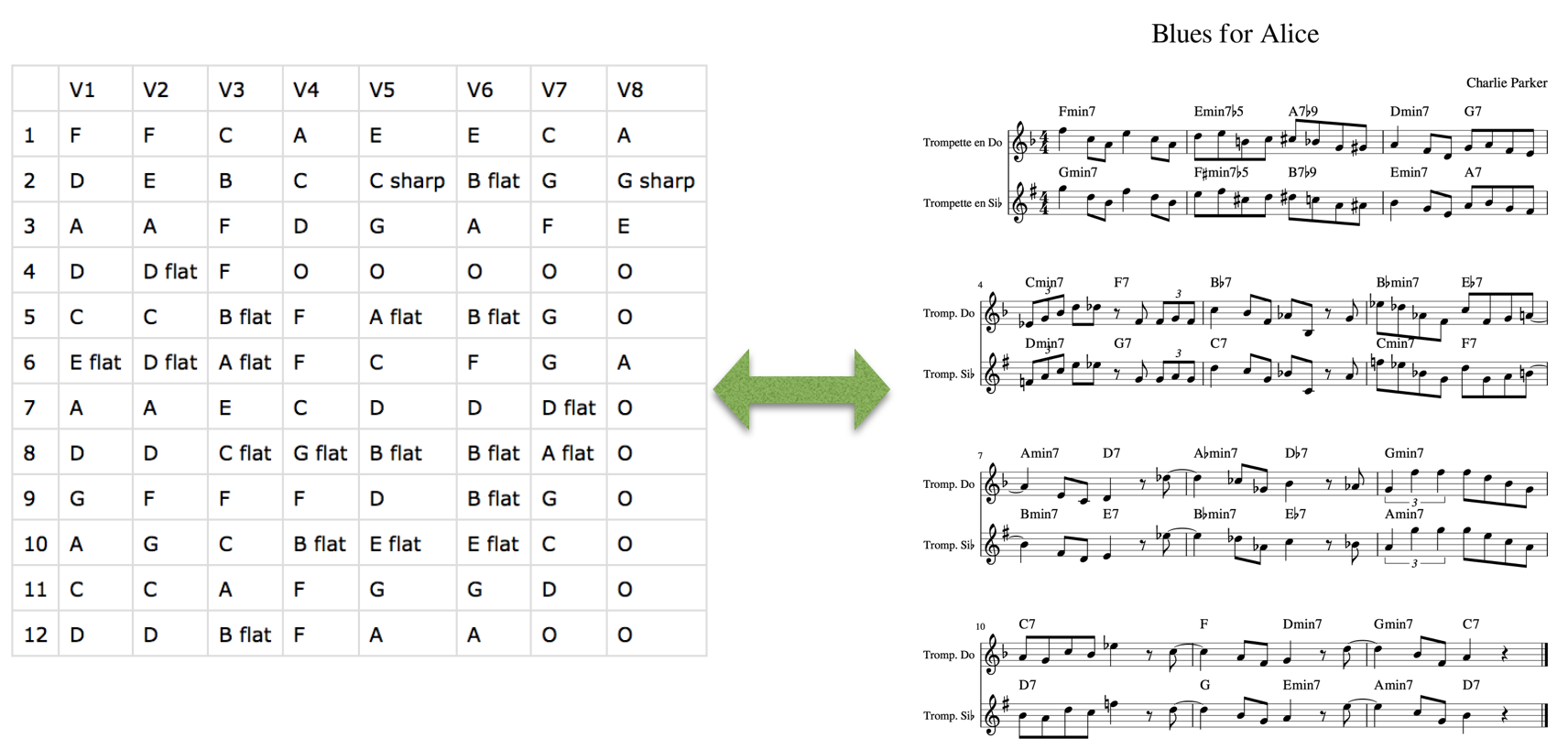}
  \caption{Transforming Notes from Music Sheets to Matrices}
 \end{figure}

Based on the concept of duration (the length of time a pitch/ tone is sounded), and in each measure the duration is fixed, we can create Measure-Note matrices. In Measure-Note matrices, we use letter \{C, D, E, F, G, A, B\} to denote the notes from "Do" to "Si", "flat" and "sharp" to denote $\flat$ and $\sharp$, and "O" to denote the rest\footnote[3]{A rest is an interval of silence in a piece of music.}.\\

As demonstrated above, for Jazz part we mainly studied work from 7 Jazz musicians (Duke Ellington, Miles Davis, John Coltrane, Charlie Parker, Louis Armstrong, Bill Evans, Thelonious Monk), and for the comparison with other music genres we focused on Chinese, Japanese, and Arabic music. So we created two different albums based on the Measure-Note matrices we generated in previous Step. I use two different ways to demonstrate the album.

\subsection{Note-Based Representation}

\begin{figure}[H]
\centering
\includegraphics[width=0.3\linewidth]{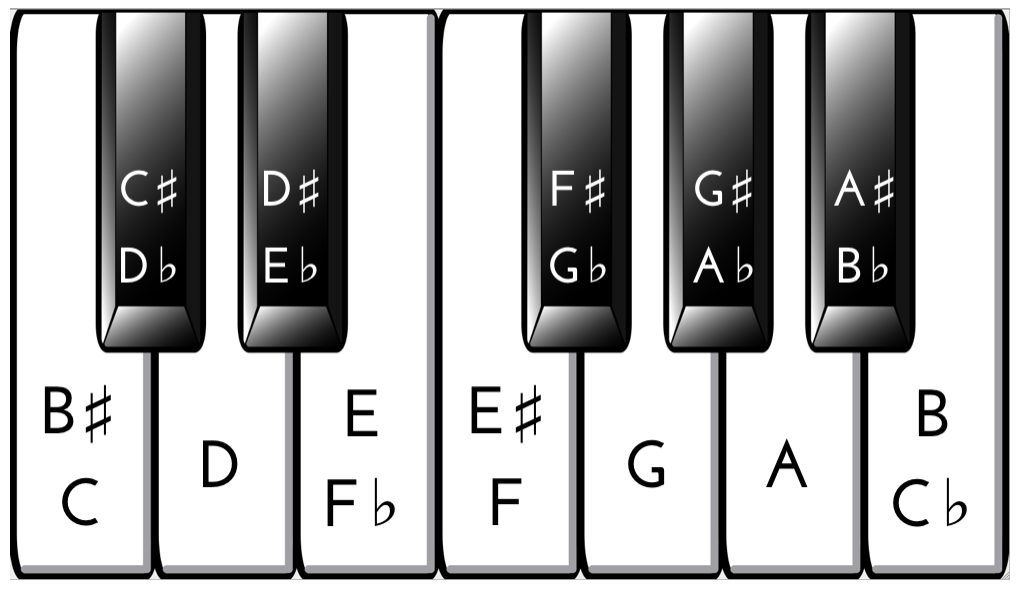}
\caption[Titanic in Music and Text]{Music Key}
\label{fig:key}
\end{figure}

Based on the 12 keys (5 black keys + 7 white keys) in the Figure \ref{fig:key}, I make note-based representation according to the pitch class in Table \ref{table:pitch}: forsaking the order of notes, we describe each measure in the song as a 12-dimension binary vector $\mathbf{X} = [x_1,x_2,...x_12]$, where $x_i\in\{0,1\}$ (Table \ref{table:2.2})

% =======
% TABLE 
% =======

\begin{table}[H]
\centering 
  \caption{Pitch Class} 
\begin{tabular}{@{\extracolsep{5pt}} ccc} 
\\[-1.8ex]\hline 
\hline \\[-1.8ex] 
Pitch Class  & Tonal Counterparts & Solfege  \\ 
\hline \\[-1.8ex]
1	& $C$, $B\sharp$	& do\\ 
2	& $C\sharp$, $D\flat$ & \\ 
3	& $D$  &	re\\ 
4   & $D\sharp$, $E\flat$  &  \\
5	& $E$, $F\flat$	& mi\\
6   & $F$, $E\sharp$ & fa\\ 
7	& $F\sharp$, $G\flat$ & \\
8	& $G$	& sol\\ 
9	& $G\sharp$, $A\flat$ & \\ 
10	& $A$ & la\\ 
11   & $A\sharp$, $B\flat$	& \\
12	& $B$, $C\flat$	& ti\\
\hline \\[-1.8ex] 
\end{tabular} \label{table:pitch}
\end{table}

% =======
% TABLE 
% =======

\begin{table}[H]
\centering 
  \caption{Notes collection from 4 Music Genres} 
\begin{tabular}{@{\extracolsep{5pt}} ccc} 
\\[-1.8ex]\hline 
\hline \\[-1.8ex] 
 Document & Pitch Class & Genre  \\ 
\hline \\[-1.8ex]
 China 1	& 0	 0	0	0	1	0	1	0	0	0	0	1	& China\\ 
China 2	& 0	0	0	0	1	0	1	0	0	0	0	0 & China\\ 
China 3	& 0	0	0	0	0	0	1	0	0	0	0	1  &	China\\ 
 \vdots &  \vdots  &  \vdots\\
China 7	& 0	1	0	0	1	0	1	0	0	0	0	1	& China\\
China 8	& 0	0	0	0	1	0	1	0	0	0	0	1	& China\\
 \vdots &  \vdots  &  \vdots \\
Japan 1	& 1	0	1	1	0	0	1	0	0	0	0	0  & Japan\\ 
Japan 2	& 1	0	0	0	0	0	0	1	0	0	0	0 & Japan\\
 \vdots &  \vdots  & \vdots\\
\hline \\[-1.8ex] 
\end{tabular}  \label{table:2.2}
\end{table} 

\begin{itemize}
\item Document: song names, tantamount to document in text mining
\item Pitch Class: binary vector whose element indicates if certain note is on, tantamount to word in text mining
\item Genre: labeled contain Chinese songs, Japanese songs, Arabic songs, to compare with Jazz songs later
\item The dimension of this data frame is $1469 \times 3$
\end{itemize}

Create the document term matrix (DTM) whose cells reflect the frequency of terms in each document. The rows of the DTM represent documents and columns represent term in the corpus. $A_{i,j}$ contains the number of times term $j$ appeared in document $i$.

% =======
% TABLE 
% =======
\begin{table}[H]
\centering 
  \caption{Document Term Matrix} 
\begin{tabular}{@{\extracolsep{5pt}} ccccc} 
\\[-1.8ex]\hline 
\hline \\[-1.8ex] 
 & Term \\ 
 Document & \footnotesize 000000000000 &	\footnotesize 000000000100 &	\footnotesize  000000010100 & ...\\
\hline \\[-1.8ex]
Arab 5	& 15	& 6 & 20 & ...\\ 
Arab 7	& 0 & 5 & 5 & ...\\ 
China 6  & 1 & 12 & 0   & ...\\
 China 7 & 13  & 0 &1  & ...\\
Japan 4 & 8 & 4  & 1  & ...\\
Japan 5  & 0 & 0  & 0  & ...\\
 USA 4  & 2 & 1  & 0 & ... \\
 \vdots &  \vdots  & \vdots & \vdots \\
\hline \\[-1.8ex] 
\end{tabular} 
\end{table}

\subsection{Measure-Based Representation}

% =======
% TABLE 
% =======
\begin{table}[H]
\centering 
  \caption{Notes collection from 7 musicians} 
\begin{tabular}{@{\extracolsep{5pt}} ccc} 
\\[-1.8ex]\hline 
\hline \\[-1.8ex] 
 Document & Notes & Musician  \\ 
\hline \\[-1.8ex]
 Charlie 1	& B$\flat$ O O O O O O O	& Charlie\\ 
 Charlie 1	& B  B$\flat$ A  A$\flat$ G  G  G$\flat$ F & Charlie \\ 
Charlie 1	& E  F  G$\flat$ B$\flat$ G  G  A$\flat$ O &	Charlie\\ 
 \vdots &  \vdots  &  \vdots\\
Charlie 7	& E  E  E  E  G  G  C  O	& Charlie\\
Charlie 8	& F$\sharp$ O O O O O O O	& Charlie\\
 \vdots &  \vdots  &  \vdots \\
Duke 1	& C  C  C  G  G  G  G  G 	& Duke\\ 
Duke 1	&F  F  F  A$\flat$ A$\flat$ A$\flat$ B$\flat$ B$\flat$	&Duke\\
 \vdots &  \vdots  & \vdots\\
\hline \\[-1.8ex] 
\end{tabular} 
\end{table} 

\begin{itemize}
\item Document: song names, tantamount to document in text mining
\item Notes: a series of notes in one measure, tantamount to word in text mining
\item Musician: the composer, tantamount to the label for later analysis
\item The dimension of this data frame is $5149 \times 3$
\end{itemize}

Create the document term matrix (DTM) whose cells reflect the frequency of terms in each document. The rows of the DTM represent documents and columns represent term in the corpus. $A_{i,j}$ contains the number of times term $j$ appeared in document $i$. Dimension of DTM is $83 \times 2960$ with the last column as label: Duke, Miles, John, Charlie, Louis, Bill, Monk.

% =======
% TABLE 
% =======
\begin{table}[H]
\centering 
  \caption{Document Term Matrix} 
\begin{tabular}{@{\extracolsep{5pt}} ccccc} 
\\[-1.8ex]\hline 
\hline \\[-1.8ex] 
 & Term \\ 
 Document & \footnotesize O O O O O O O O &	\footnotesize B  D  B  B  D  D  E  E &	\footnotesize  C  A  A$\sharp$ B  D  C  A  O	& ...\\
\hline \\[-1.8ex]
Miles 6	& 40	& 0 & 0 & ...\\ 
Louis 2	& 32 & 0 & 0 & ...\\ 
Sonny 3 &  26 &  0 &  0 & ...\\
Miles 2	& 25	& 0& 0 & ...\\
Duke 4	& 0& 9& 0 & ...\\
Sonny 4 &  14  &  0 & 0& ... \\
Charlie 9 & 0	& 0 & 8& ...\\
 \vdots &  \vdots  & \vdots & \vdots \\
\hline \\[-1.8ex] 
\end{tabular} 
\end{table} 

We can also talk a close look at the most frequent terms in the whole album: terms appear more than 20 times:
% =======
% TABLE 
% =======
\begin{table}[H]
\centering 
  \caption{Most Frequent Terms} 
\begin{tabular}{@{\extracolsep{5pt}} ccc} 
\\[-1.8ex]\hline 
\hline \\[-1.8ex] 
& Term & \\
\hline \\[-1.8ex]
&O O O O O O O O\\
&C  C  C  C  C  C  C  C\\
&A  A  A  A  O  O  O  O\\
&B$\flat$ B$\flat$ B$\flat$ B$\flat$ B$\flat$ B$\flat$ B$\flat$ B$\flat$\\ 
&B  B  B  B  B  B  B  B\\
&D  D  D  D  D  D  D  D\\
&G  G  G  G  G  G  G  G\\
&A  A  A  A  A  A  A  A \\
\hline \\[-1.8ex] 
\end{tabular} 
\end{table}

\section{Pattern Recognition}

We take the topic proportion matrix as input and employ it on machine learning techniques for classification. We conduct the supervised analysis via 5 models with k-fold cross-validation:
\begin{itemize}
\item K Nearest Neighbors
\item Multi-class Support Vector Machine 
\item Random Forest
\item Neural Networks with PCA Analysis
\item Penalized Discriminant Analysis
\end{itemize}

% =======
% ALGORITHM. 
% =======
\begin{algorithm}[H]
\caption{Supervised Analysis: 10-fold cross-validation with 3 times resampling}
\begin{algorithmic} 
\FOR{$i \gets 1:3$}
	\FOR{$j \gets 1:10$}
	\STATE Split dataset $\mathcal{D} = \{ \mathbf{z}_l, l = 1,2,...,n\}$ into $k$ chunks so that $n = Km$
	\STATE Form subset $\mathcal{V}_j = \{\mathbf{z}_l\in \mathcal{D}: i\in [1+(j-1)\times m, j\times m]\}$
	\STATE Extract train set $\mathcal{T}_j:= \mathcal{D}\backslash \{\mathcal{V}_j\}$
	\STATE Build estimator $\hat{g}^{(\star)}(\cdot)$ using $\mathcal{T}_j$
	\STATE Compute predictions $\hat{g}^{(j)}(\mathbf{x}_l)$ for $\mathbf{z}_k\in \mathcal{V}_j$
	\STATE Calculate the error $\hat{\epsilon}_j = \frac{1}{m}\sum_{\mathbf{z}_l\in \mathcal{V}_j} l(y_l, \hat{g}^{(j)}(\mathbf{x}_l))$
	\ENDFOR \\
\STATE Compute $\texttt{CV}(\hat{g}) = \frac{1}{K}\sum_{j=1}^K \hat{\epsilon}_j$
\STATE Find $\hat{g}^{(\star)}(\cdot) = \underset{j=1:J}{\texttt{argmin}}\{\texttt{CV}(\hat{g}(\cdot))\}$ with lowest prediction error
\ENDFOR
\end{algorithmic}
\end{algorithm}

\subsection{K-Nearest Neighbors}
kNN predicts the class of song via finding the k most similar songs, where the similarity is measured by Euclidean distance between two song vectors in this case. The class (label) here is the 7 musicians: Duke, Miles, John, Charlie, Louis, Bill, Monk.   

% =======
% ALGORITHM. 
% =======
\begin{algorithm}[H]
\caption{k-Nearest Neighbors}
\begin{algorithmic} 
\FOR{$i \gets 1:n$}
	\STATE Choose the value of $k$ for $\mathcal{D} = \{(\mathbf{x}_1, Y_1), ...,(\mathbf{x}_i, Y_i), ..., (\mathbf{x}_n, Y_n), Y_i\in \{1,...,g\}\}$
	\STATE Let $\mathbf{x}^\star$ be a new point. Compute $d_i^\star = d(\mathbf{x}^\star, \mathbf{x}_i)$
\ENDFOR \\
\STATE Rank all the distance $d_i^\star$ in order: $d_{(1)}^\star\leq d_{(2)}^\star \leq ...\leq d_{(k)}^\star \leq... \leq d_{(n)}^\star$
\STATE Form $\mathcal{V}_k(\mathbf{x}^\star) = \{\mathbf{x}_i:d(\mathbf{x}^\star, \mathbf{x}_i)\leq  d_{(k)}^\star  \}$
\STATE Predict response $\hat{Y}^\star_{kNN} = \text{Most frequent label in }\mathcal{V}_k(\mathbf{x}^\star) = \underset{j\in \{1,...,g\}}{\texttt{argmax}} \{p_j^{(k)}(\mathbf{x}^\star)\}$
\STATE where $p_j^{(k)}(\mathbf{x}^\star) = \frac{1}{k}\sum_{\mathbf{x}_i\in \mathcal{V}_k(\mathbf{x}^\star)}\mathbf{I}(Y_i = j) $ 
\end{algorithmic}
\end{algorithm}

\subsection{Support Vector Machine}
The task of Support Vector Machine (SVM) is to find the \textit{optimal hyperplane that separates the observations in such a way that the margin is as large as possible.} That is to say, the distance between the nearest sample patterns (support vectors) should be as large as possible. SVM is originally designed as binary classifier, so in this case there are more than two classes, we use multi-class SVM. Specifically, we transform single multi-class task into multiple binary classification task. We train $K$ binary SVMs and maximize the margins from each class to the remaining ones. We choose linear kernel (Eq.\ref{linearkernal})   due to its excellent performance on high dimensional data that are very sparse in text mining. 
\begin{align}
\mathcal{K}(\mathbf{x}_i, \mathbf{x}_j) = < \mathbf{x}_i, \mathbf{x}_j> = \mathbf{x}_i^\top \mathbf{x}_j \label{linearkernal}
\end{align}
% =======
% ALGORITHM. 
% =======
\begin{algorithm}[H]
\caption{Multi-class Support Vector Machine}
\begin{algorithmic} 
\FOR{$k \gets 1:K$}
	\STATE Given $\mathcal{D} = \{(\mathbf{x}_1, Y_{1k}), ...,(\mathbf{x}_i, Y_{ik}), ..., (\mathbf{x}_n, Y_{nk}), Y_{ik}\in \{+1,-1\}\}$
	\STATE Find function $h(\mathbf{x}) = \mathbf{w}^\top\mathbf{x}+b$ that achieves
	\STATE $\underset{\mathbf{w},b}{\text{max}}\left[ \underset{y_{ik} = +1}{\text{min}}\left( \frac{|\mathbf{w}^\top\mathbf{x}_i+b|    }{||\mathbf{w}||} \right) + \underset{y_{ik} = -1}{\text{min}}\left( \frac{|\mathbf{w}^\top\mathbf{x}_i+b|    }{||\mathbf{w}||} \right)      \right] = \underset{\mathbf{w},b}{\text{max}} \frac{2}{||\mathbf{w}||} = \underset{\mathbf{w},b}{\text{min}} \frac{1}{2}||\mathbf{w}||^2   $
	\STATE subject to $Y_{ik}(\mathbf{w}^\top\mathbf{x}_i + b)\geq 1, \forall i = 1,2,...,n$
\ENDFOR \\
Get $\underset{k=1,...,K}{\texttt{argmax} } f_k(\mathbf{x}) = \underset{k=1,...,K}{\texttt{argmax} } (  \mathbf{w}_k^\top\mathbf{x}+b_k) $
\end{algorithmic}
\end{algorithm}

\subsection{Random Forest}

Random Forest (RF) as an ensemble learning method that optimal the performance of single tree. Compared with tree bagging, the only difference in random forest is that then select each tree candidate with random subset of features, called \textit{"feature bagging"}, for correction of overfitting issue of trees. If some features weigh more strongly than other features, these features will be selected in many of $B$ trees among the whole forest. 

% =======
% ALGORITHM. 
% =======
\begin{algorithm}[H]
\caption{Random Forest}
\begin{algorithmic} 
\FOR{$b \gets 1:B$}
	\STATE Draw with replacement from $\mathcal{D}$ a sample $\mathcal{D}^{(b)} = \{\mathbf{z}_1^{(b)},..., \mathbf{z}_n^{(b)} \}$ 
	\STATE Draw subset $\{i_1^{(b)}, ..., i_d^{(b)}\}$ of $d$ variables without replacement from $\{1,2,...,p\}$ 
	\STATE Prune unselected variables from the sample $\mathcal{D}^{(b)}$ to ensure $\mathcal{D}^{(b)}_{sub}$ is $d$ dimension
	\STATE Build tree (base learner) $\hat{g}_{(b)}$ based on $\mathcal{D}^{(b)}_{sub}$ 
\ENDFOR \\
Output the result based on the mode of classes $ \hat{g}^{RF}(\mathbf{x})=  \underset{j\in \{1,...,B\}}{\texttt{argmax}} \{p_j^{(b)}(\mathbf{x})\} $ 
\STATE where $p_j^{(k)}(\mathbf{x}^\star) = \frac{1}{B}\sum \mathbf{I}(\hat{g}_{(b)}(\mathbf{x})= j) $ 
\end{algorithmic}
\end{algorithm}

\subsection{Neural Network with PCA Analysis}

Principal Components Analysis (PCA) as one of the most common dimension reduction methods can help improve the result of classification. Neural Network with Principal Component Analysis method proposed by \citet{2007pattern} is to run principal component analysis on the data first and then use the component in the neural network model. Each predictor has more than one values as the variance of each predictor is used in PCA analysis, and the predictor only has one value would be removed before the analysis. New data for prediction are also transformed with PCA analysis before feed to the networks.

% =======
% ALGORITHM. 
% =======
\begin{algorithm}[H]
\caption{Neural Network with PCA Analysis}
\begin{algorithmic} 
\STATE Given data $\mathcal{D} = \{\mathbf{x}_1,..., \mathbf{x}_n \}, \mathbf{x}_i \in \mathbb{R}^m$, finding $\hat{\Sigma}$ as estimates 
\FOR{$i \gets 1:p$}
\STATE Obtain eigenvalues $\hat{\lambda}_i$ and eigenvectors $\hat{e}_i$ from $\hat{\Sigma}$
\STATE Obtain principal components $y_i = \hat{e}_j^\top X$
\ENDFOR \\
	\STATE Get $p$-dimensional input vector $\mathbf{y} = (y_1,y_2,...,y_p)^\top$ after PCA analysis 
\FOR{$j \gets 1:q$}
	\STATE Compute linear combination $h_j(\mathbf{y}) = \beta_{0j} + \mathbf{\beta}_j^\top\mathbf{y}$ for each node in hidden layer 
	\STATE Pass $h_j(\mathbf{y})$ through nonlinear activation function $z_j = \psi(\beta_{0j} + \sum_{l=1}^p\beta_{lj}y_l )$
\ENDFOR \\
	\STATE Combine $z_j$ with coefficients to get $\eta(\mathbf{y}) = \gamma_0 + \sum_{j=1}^q \gamma_j  \psi(\beta_{0j} + \sum_{l=1}^p\beta_{lj}y_l ) $
	\STATE Pass $\eta(\mathbf{y})$ with another activation function to output layer $\mu_k(\mathbf{y}) = \phi_k(\eta(\mathbf{y}) )$
\end{algorithmic}
\end{algorithm}

\subsection{Penalized Discriminant Analysis}

Linear Discriminant Analysis (LDA) is common tool for classification and dimension reduction. However, LDA can be too flexible in the choice of $\mathbf{\beta}$ with highly correlated predictor variables.  \citet{1995penalized} came up with Penalized Discriminant Analysis (PDA) to avoid the overfitting performance resulting from LDA. Basically a penalty term is added to the covariance matrix $\Sigma_W' = \Sigma_W + \Omega$.

% =======
% ALGORITHM. 
% =======
\begin{algorithm}[H]
\caption{Penalized Discriminant Analysis}
\begin{algorithmic} 
\FOR{$i \gets 1:n$} 
	\STATE Given data $\mathcal{D} = \{(\mathbf{x}_1,Y_1),..., (\mathbf{x}_n, Y_n)\}, \mathbf{x}_i \in \mathbb{R}^q$ 
	\STATE Compute within-class covariance matrix $\hat{\Sigma}_w =\sum_{i=1}^n(\mathbf{x}_i- \mathbf{\mu}_{y_i})(\mathbf{x}_i- \mathbf{\mu}_{y_i})^\top +\Omega$
	\STATE Compute between-class covariance matrix $\hat{\Sigma}_b = \sum_{j=1}^m n_j(\mathbf{x}_j- \mathbf{\mu}_{y_j})(\mathbf{x}_j - \mathbf{\mu}_{y_j})^\top $
\ENDFOR \\
	\STATE Maximize the ratio of two matrices: $\hat{\mathbf{w}} = \underset{\mathbf{w}}{\texttt{argmax}} \frac{\mathbf{w}^\top \hat{\Sigma}_b \mathbf{w}}{\mathbf{w}^\top \hat{\Sigma}_w \mathbf{w}}$
	%\STATE Prior probability of a selected entity from class is $j$ is $\pi_j = Pr[Y_i=j]$ 
	%\STATE Conditional density $p(\mathbf{x}|y=j)=\frac{1}{(2\pi)^{q/2} |\Sigma_j|^{1/2}}\text{exp}\left\{-\frac{1}{2}(\mathbf{x}-\mathbf{\mu}_j)^\top  \Sigma_j^{-1} (\mathbf{x}-\mathbf{\mu}_j) \right\} $
	%\STATE Compute discriminant function $\delta_j(\mathbf{x}_i) = Pr[Y_i=j|\mathbf{x}] = \frac{\pi_jp(\mathbf{x}|y=j)}{p(\mathbf{x})}$
	%\STATE Assign $\mathbf{x}_i$ to the class with the highest posterior probability
\end{algorithmic}
\end{algorithm}

\section{Topic Modeling}

\subsection{Intuition Behind Model}
\begin{figure}[H]
 \centering
  \includegraphics[scale=0.18]{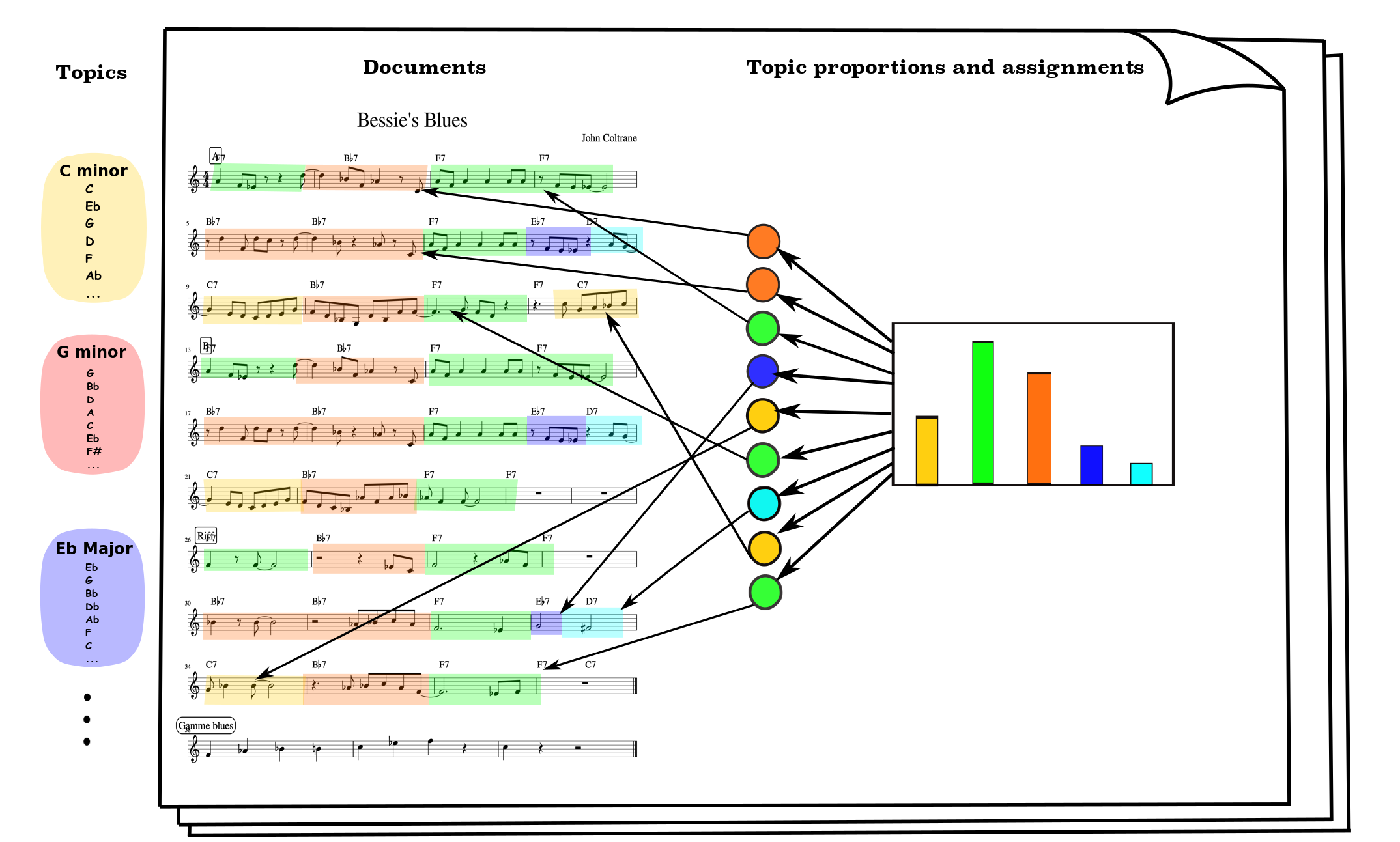}
  \caption{Intuition behind Music Mining}\label{music}
 \end{figure}

Similar to the work from \citet{blei2012} in text mining, Figure \ref{music} illustrates the intuition behind our model in music concept. We assume an album, as a collection of songs, are mixture of different topics (melodies). These topics are the distributions over a series of notes (left part of the figure). In each song, notes in every measure are chosen based on the topic assignments (colorful tokens), while the topic assignments are drawn from the document-topic distribution. 

\subsection{Model}

\begin{figure}[H]
  \centering
  \tikz{ %
    \node[latent] (alpha) {$\alpha$} ; %
    \node[latent, right=of alpha] (theta) {$\theta$} ; %
    \node[latent, right=of theta] (z) {z} ; %
    \node[obs, right=of z] (u) {u} ; %
    \node[latent, right=of u] (beta) {$\beta$} ; %
    \node[latent, right=of beta] (eta) {$\eta$} ; %
    \plate[inner sep=0.25cm, xshift=-0.12cm, yshift=0.10cm] {plate4} {(u)} {L}; %
    \plate[inner sep=0.30cm, xshift=-0.12cm, yshift=0.12cm] {plate1} {(z) (u)} {\shft{N}}; %
    \plate[inner sep=0.30cm, xshift=-0.12cm, yshift=0.12cm] {plate2} {(theta) (plate1)} {M}; %
    \plate[inner sep=0.25cm, xshift=0cm, yshift=0.12cm] {plate3} {(beta)} {K}; %
    \edge {alpha} {theta} ; %
    \edge {theta} {z} ; %
    \edge {z,beta} {u} ; %
    \edge {eta} {beta}; 
  }
\end{figure}

\begin{align}
\textbf{Dirichlet: }\quad p(\theta|\alpha) &= \frac{\Gamma(\sum_i\alpha_i)}{\prod_i\Gamma(\alpha_i)}\prod_{i=1}^K\theta_i^{\alpha_i-1}\quad 
& p(\beta|\eta) = \frac{\Gamma(\sum_i\eta_i)}{\prod_i\Gamma(\eta_i)}\prod_{i=1}^K\theta_i^{\eta_i-1}\\ 
\textbf{Multinomial: }\quad p(z_n|\theta) &= \prod_{i=1}^K\theta_i^{z_n^i}\quad
 &p(x_n|z_n,\beta) = \prod_{i=1}^K \prod_{j=1}^V\beta_{ij}^{(z_n^ix_n^j)}
\end{align}

\subsubsection*{Notation}
\begin{itemize}
\item $u$: notes (observed)
\item $z$: chord per measure (hidden)
\item $\theta$ chord proportions for a song (hidden)
\item $\alpha$: parameter controls chord proportions 
\item $\beta$: key profiles
\item $\eta$: parameter controls key profiles
\end{itemize}

\subsection{Generative Process}
\begin{enumerate}
\item Draw $\theta \sim \text{Dirichlet}(\alpha)$
\item For each harmony $k \in \{1,...,K\}$
\begin{itemize}
\item Draw $\beta_k\sim \text{Dirichlet}(\eta)$
\end{itemize}
\item For each measure $\textbf{u}_n$ (notes in nth measure) in song $m$
\begin{itemize}
\item Draw harmony $z_n\sim \text{Multinomial}(\theta)$
\item Draw pitch in nth measure $x_n|z_n\sim \text{Multinomial}(\beta_{k})$
\end{itemize}
\end{enumerate}

\subsubsection*{Terms for single song:}

\begin{align}
p(\theta|\alpha) &= \frac{\Gamma(\sum_i\alpha_i)}{\prod_i\Gamma(\alpha_i)}\prod_{i=1}^K\theta_i^{\alpha_i-1}\\
p(\beta|\eta) &= \frac{\Gamma(\sum_i\eta_i)}{\prod_i\Gamma(\eta_i)}\prod_{i=1}^K\theta_i^{\eta_i-1}\\ 
p(z_n|\theta) &= \prod_{i=1}^K\theta_i^{z_n^i}\\
p(x_n|z_n,\beta) &= \prod_{i=1}^K \prod_{j=1}^V\beta_{ij}^{(z_n^ix_n^j)}
\end{align}
\subsubsection*{Joint Distribution for the whole album:}
\begin{align}
p(\theta,\mathbf{z},\mathbf{x}|\alpha,\beta,\eta) &= \prod_{k=1}^K p(\beta|\eta)\prod_{m=1}^M p(\theta|\alpha)\Big(\prod_{n=1}^N p(z_n|\theta) p(x_n|z_n,\beta)\Big)
\end{align}

\subsubsection*{Summary}
\begin{itemize}
\item Assume there are M documents in the corpus.
\item The topic distribution under each document is a Multinomial distribution $Mult(\theta)$ with its conjugate prior $Dir(\alpha)$.
\item The word distribution under each topic is a Multinomial distribution $Mult(\beta)$ with the conjugate prior $Dir(\eta)$.
\item For the $n^{th}$ word in the certain document, first we select a topic $z$ from per document-topic distribution $Mult(\theta)$, then select a word under this topic $x|z$ from per topic-word distribution $Mult(\beta)$.
\item Repeat for M documents. For M documents, there are M independent Dirichlet-Multinomial Distributions; for K topics, there are K independent Dirichlet-Multinomial Distributions.
\end{itemize}

\subsection{Estimation}

For per-document posterior is 
\begin{align}
p(\beta, \mathbf{z},\theta|\mathbf{x}, \alpha, \eta) &= \frac{p(\theta, \beta, \mathbf{z}, \mathbf{x}|\alpha, \eta)}{p(\mathbf{x}|\alpha,\eta)} = \frac{p(\theta|\alpha)\prod_{n=1}^N p(z_n|\theta)p(x_n|z_n, \beta_{1:K}) }{\int_{\theta} p(\theta|\alpha)\prod_{n=1}^N \sum_{z=1}^K p(z_n|\theta)p(x_n|z_n, \beta_{1:K})  }
\end{align}
Here we use Variational EM (VEM) instead of EM algorithm to approximate posterior inference because the posterior in E-step is intractable to compute. 

 \begin{figure}[H]
  \centering
  \includegraphics[scale=0.4]{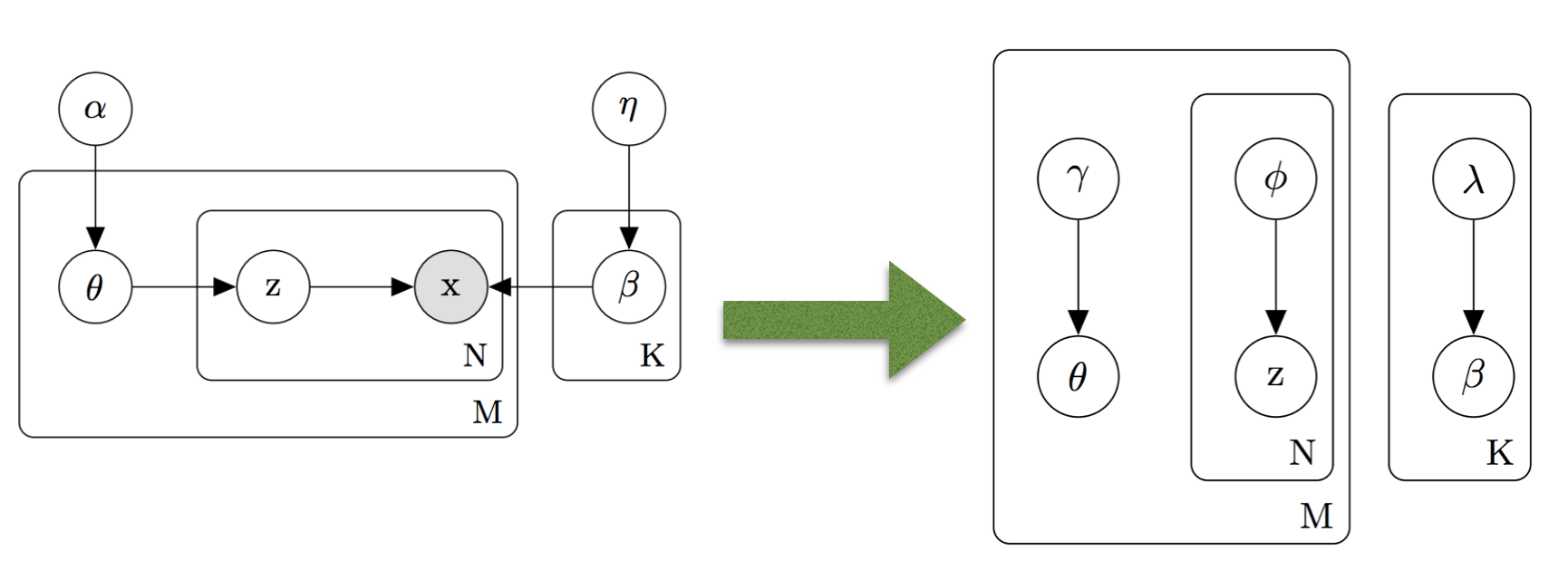}
    \caption{Variational EM Graphical Model}\label{VEM}
 \end{figure}

\citet{blei2003} proposed a way to use variational term $q(\beta, \mathbf{z},\theta|\lambda, \phi, \gamma) $ (Eq.\ref{q}) to approximate the posterior $p(\beta, \mathbf{z},\theta|\mathbf{x}, \alpha, \eta) $ (Eq.\ref{p}). That is to say, by removing certain connections in the graphical model in Figure \ref{VEM}, we obtain the tractable version of lower bounds on the log likelihood.

\begin{align}
q(\beta, \mathbf{z},\theta|\lambda, \phi, \gamma) &= \sum_{k=1}^K \text{Dir}(\beta_k|\lambda_k)\sum_{d=1}^M (q(\theta_d|\gamma_d) \sum_{n=1}^N q(z_{dn}|\phi_{dn}) ) \label{q}\\
p(\beta, \mathbf{z},\theta|\mathbf{x}, \alpha, \eta) &= \frac{p(\theta, \beta, \mathbf{z}, \mathbf{x}|\alpha, \eta)}{p(\mathbf{x}|\alpha,\eta)}\label{p}
\end{align}
With the simplified version of posterior distribution, we aim to minimize the KL Distance (Kullback–Leibler divergence) between the variational distribution $q(\beta, z,\theta|\lambda, \phi, \gamma) $  and the posterior  $p(\beta, z,\theta|x, \alpha, \eta) $ to obtain the optimal value of the variational parameters $\gamma$, $\phi$, and $\lambda$ (Eq.\ref{kb}). That is to obtain the maximum lower bound $L(\gamma,\phi,\lambda;\alpha, \eta)$ (Eq.\ref{lowerbound}). 

\begin{align}
\text{ln}p(\mathbf{x|\alpha, \eta})&= L(\gamma,\phi,\lambda;\alpha, \eta)  + D(q(\beta, \mathbf{z},\theta|\lambda, \phi, \gamma) ||p(\beta, \mathbf{z},\theta|\mathbf{x}, \alpha, \eta)) \\
(\lambda^*,\phi^*, \gamma^*) &= \underset{\lambda, \phi,\gamma}{\texttt{argmin}}D(q(\beta, \mathbf{z},\theta|\lambda, \phi, \gamma) ||p(\beta, \mathbf{z},\theta|\mathbf{x}, \alpha, \eta)) \label{kb}\\ 
L(\gamma,\phi,\lambda;\alpha, \eta) &= E_q[\text{ln}p(\theta | \alpha)] +  E_q[\text{ln}p(\mathbf{z} | \theta)] + E_q[\text{ln}p(\beta | \eta)] + E_q[\text{ln}p(\mathbf{x} | \mathbf{z},\beta)] \cr 
& - E_z[\text{ln}q( \theta|\gamma)]  -E_q[\text{ln}q(\mathbf{z} | \phi)]- E_z[\text{ln}q( \beta|\lambda)] \label{lowerbound}
\end{align}

% =======
% ALGORITHM. 
% =======
\begin{algorithm}[H]
\caption{Variational EM for Smoothed LDA in Sheet Music}
\begin{algorithmic} 
\FOR{$t \gets 1:T$}
\STATE \textbf{E-step}
\STATE Fix model parameters $\alpha$, $\eta$. Initialize $\phi_{ni}^0:= \frac{1}{k}, \gamma_{i}^0:= \alpha_i+\frac{N}{k}, \lambda_{ij}^0 := \eta $
\FOR{$n \gets 1:N$}
  \FOR{$i \gets 1:k$}
    \STATE $\phi_{ni}^{t+1}:= \text{exp}(\Psi(\gamma^t_i)) \prod_{j=1}^V\beta_{ij}^{x_n^j} $
  \ENDFOR \\
  \STATE Normalize $\phi_{n}^{t+1}$ to sum to 1
\ENDFOR \\
\STATE $\gamma^{t+1}:= \alpha +\sum_{n=1}^N\phi_n^{t+1}$
\STATE $\lambda_{j}^{t+1} := \eta + \sum_{d=1}^M\sum_{n=1}^{N_d} \phi^{t+1}_{dn}x^j_{dn}$
\STATE \textbf{M-step}
\STATE Fix the variational parameters $\gamma$, $\phi$, $\lambda$
\STATE Maximize lower bound with respect to model parameters $\eta$, $\alpha$
\STATE \textbf{until converge}
\ENDFOR \\
\end{algorithmic}
\end{algorithm}

\section{Implementation}
In this section we implement pattern recognition and topic modeling methods with two representation (note-based representation and measure-based representation) demonstrated previously, and evaluate performance of different representations in diverse scenarios.  

\subsection{Pattern Recognition}
\subsubsection{Note-Based Model}
 \begin{figure}[H]
 \centering
  \includegraphics[scale=0.4]{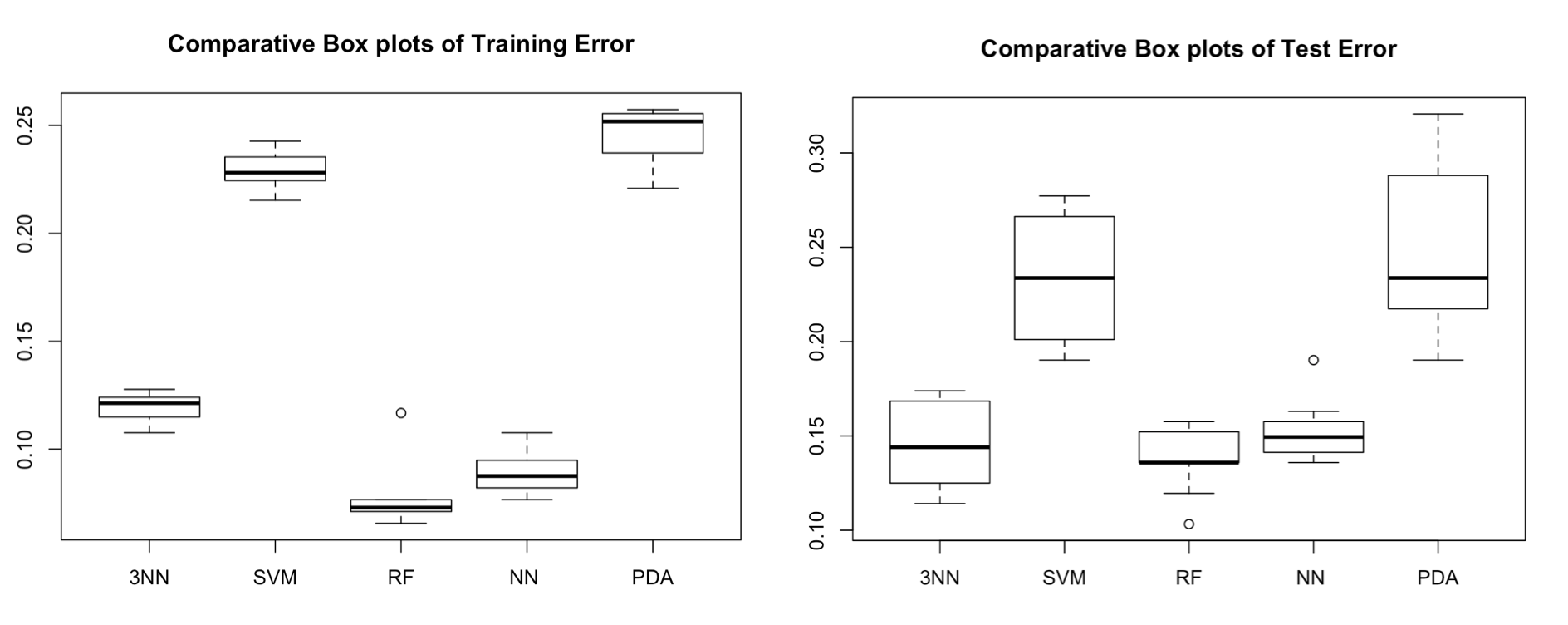}
  \caption{Pattern Recognition on Jazz and Chinese Music}
 \end{figure}

 \begin{figure}[H]
 \centering
  \includegraphics[scale=0.4]{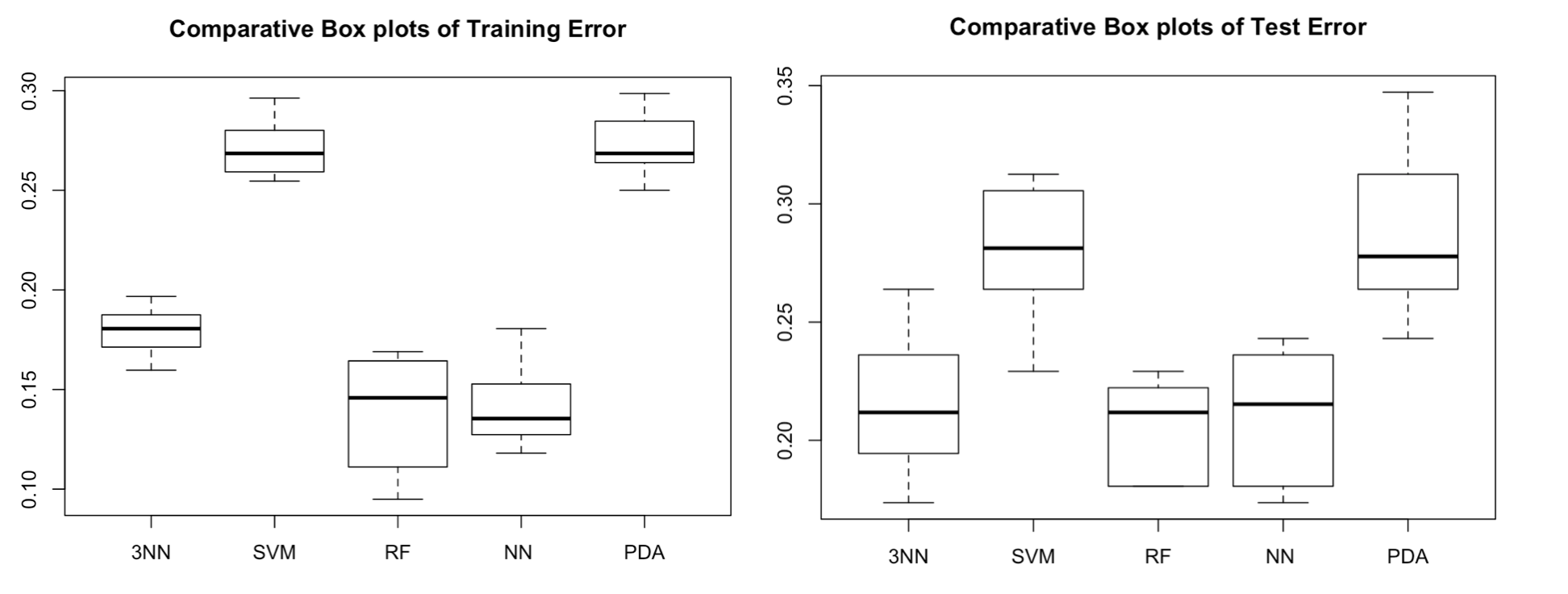}
  \caption{Pattern Recognition on Jazz and Japanese Music}
 \end{figure}

 \begin{figure}[H]
 \centering
  \includegraphics[scale=0.4]{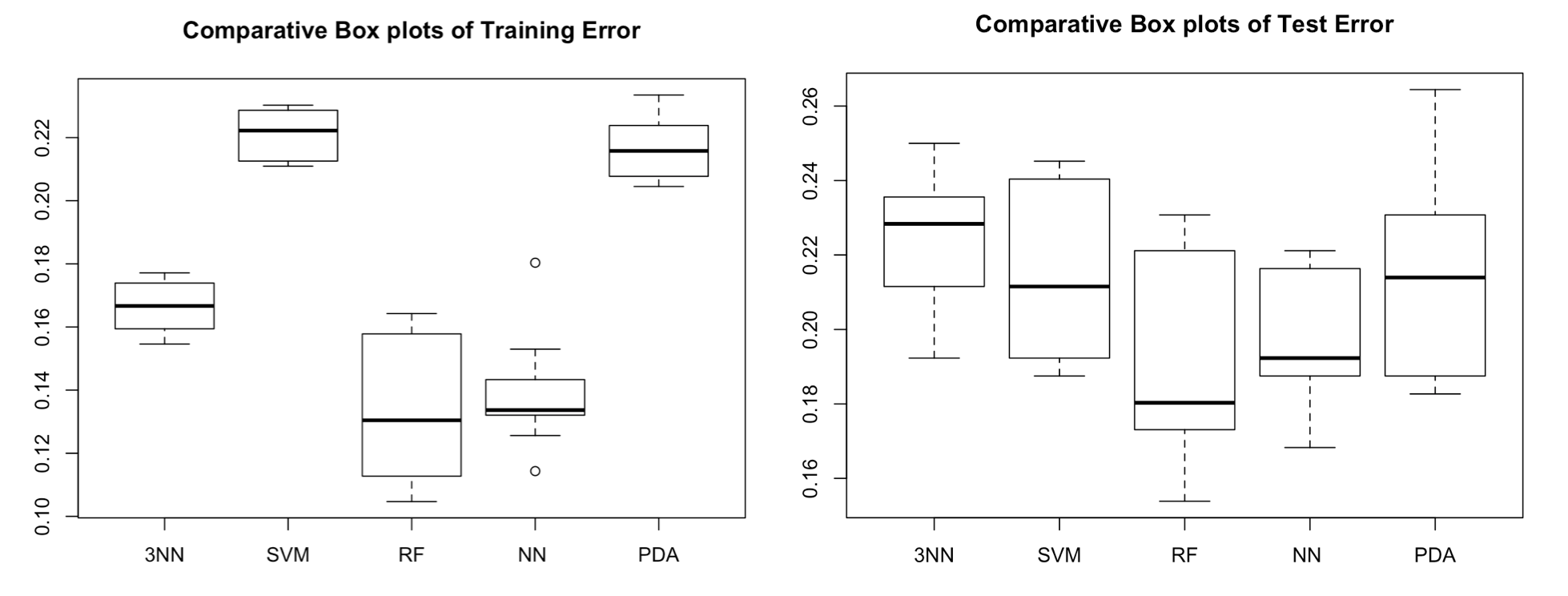}
  \caption{Pattern Recognition on Jazz and Arabic Music}
 \end{figure}

\subsubsection{Measure-Based Model}

\begin{figure}[H]
 \centering
  \includegraphics[scale=0.4]{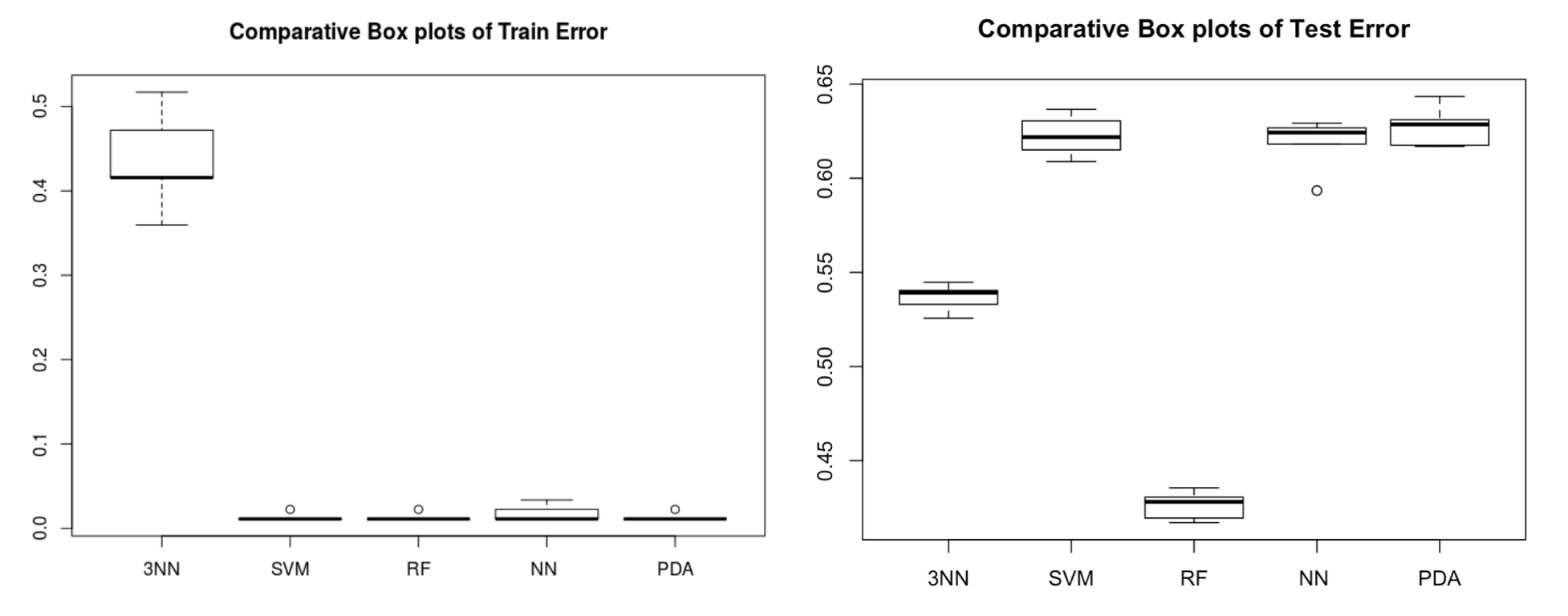}
  \caption{Pattern Recognition on Different Jazz Musicians}
 \end{figure}

\subsubsection{Comments and Conclusion}
For note-based model we can see that the five supervised machine learning techniques could all classify different music genre with error rate no more than 35\%. In addition, the performance of random forest, k nearest neighbors, and neural networks with PCA analysis are much better than the other two methods. Among the three comparisons (Jazz vs. Chinese music, Jazz vs. Japanese music, Jazz vs. Arabic music), the comparison of Jazz vs. Chinese would give better result than the other two, with random forest reaching lower than 0.1 error rate. For recognition between Jazz and Chinese songs, random forest is the best one with lowest error rate and variance. For recognition between Jazz and Japanese songs,  k nearest neighbors, neural network and random forest have comparatively low error rate, but k nearest neighbors' performance has smaller variance. For comparison between Jazz and Arabic songs, neural network and random forest have comparatively low error rate, while they all have large variance. \\ 

For measure-based model, we can see that from the confusion matrix of training set, the model accuracy rate is very high for all techniques expect k nearest neighbors. However, but for the test set all the model fails to provide very good result with lowest error rate as 0.4 from random forest. It is obvious that this scenario has the challenging of overfitting issue. Further investigation is necessary if we want to use this representation.\\

\subsection{Topic Modeling}

\subsubsection{Perplexity}

In topic modeling, the number of topics is crucial for the to achieve its optimal performance. Perplexity is one way to measure how well is predictive ability of a probability model. Having the optimal topic number is always helpful in the sense to reach the best result with minimum computational time. Perplexity of a corpus $\mathcal{D}$ of $M$ documents is computed as below Equation (\ref{eq:perplex}).

\begin{align}
P(\mathcal{D}) = \text{exp}\left(\frac{-\sum_{d=0}^{M-1}\text{log } p(w_d;\lambda )}{\sum_{d=0}^{M-1}N_d}\right) \label{eq:perplex}
\end{align}

Apart from the above common way, there are many other methods to find the optimal topics. The existing \texttt{ldatuning} package stores 4 methods to calculate all metrics for selecting the perfect number of topics for LDA model all at once. \\

Table \ref{table:tuning} shows 4 different evaluating matrices. The extrema in each scenario illustrates the optimal number of topics.
\begin{itemize}
\item minimum
\begin{itemize}
\item Arun2010 \citep{arun2010}
\item CaoJuan2009 \citep{cao2009}
\end{itemize}
\item Maximum
\begin{itemize}
\item Deveaud2014 \citep{deveaud2014}
\item Griffiths2004 \citep{griffiths2004}
\end{itemize}
\end{itemize}
% =======;
% TABLE 
% =======
\begin{table}[H]\centering 
  \caption{Perplexity of Different Matrices} \label{table:tuning}
\begin{tabular}{@{\extracolsep{5pt}} cccccccc} 
\\[-1.8ex]\hline 
\hline \\[-1.8ex] 
Topics Number & Griffiths2004 & CaoJuan2009 & Arun2010 & Deveaud2014 \\
\hline \\[-1.8ex]
2   &  -7454.086 & 0.11290217 & 13.856421  & 1.8604276\\
4    & -6821.928  & 0.07120480 & 8.508257   &1.7877936 \\
6    & -6516.431  & 0.06146701 & 5.613616   &1.7126743 \\
8    & -6322.309  & 0.05740186 & 3.728195   &1.6422201\\
10   &  -6184.650 & 0.05336498 & 2.404497   &1.5998098\\
16   &  -6112.754 & 0.06507096 & 1.328469   &1.3594688\\
20   &  -6101.264 & 0.07099931 & 1.512142   &1.2242214\\
26   &  -6129.508 & 0.09352393 & 1.856783   &1.0760613 \\
30   &  -6121.120 & 0.10582645 & 2.545512   &0.9585189\\
36   &  -6177.121 & 0.12330036 & 4.078891   &0.8530592\\
40   &  -6183.168 & 0.14128330 & 5.226102   &0.7767756\\
46   &  -6224.206 & 0.15072742 & 5.372056   &0.7119278\\
50   &  -6253.992 & 0.16448002 & 6.637710   & 0.6719547\\
60   &  -6352.595 & 0.20606817 & 7.769699  & 0.5844223\\
72   &  -6325.653 & 0.25947947 & 9.892807  & 0.4742397\\
80   &  -6393.940 & 0.26968788 & 10.187645 &  0.4463054\\
\hline \\[-1.8ex] 
\end{tabular} 
\end{table} 
\FloatBarrier

 \begin{figure}[H]
 \centering
  \includegraphics[scale=0.3]{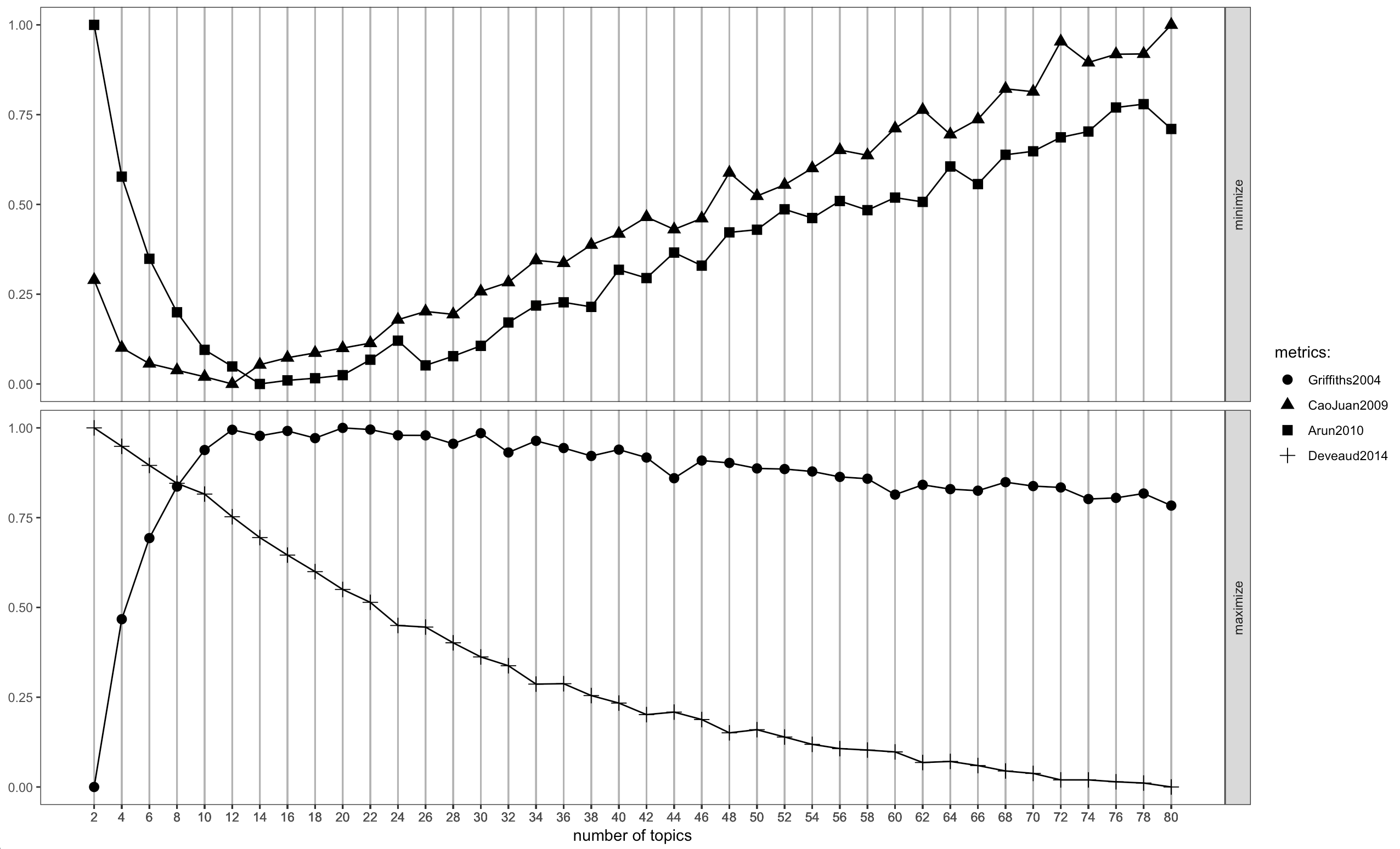}
  \caption{Evaluating LDA Models }\label{perplexity}
 \end{figure}

From perplexity we can come to the conclusion that the optimal number of topics is around 8$\sim$12. In this scenario Metric \textit{Deveaud2014} is not as informative as the other three.

\subsubsection{Discussion}
\begin{figure}[H]
 \centering
  \includegraphics[scale=0.5]{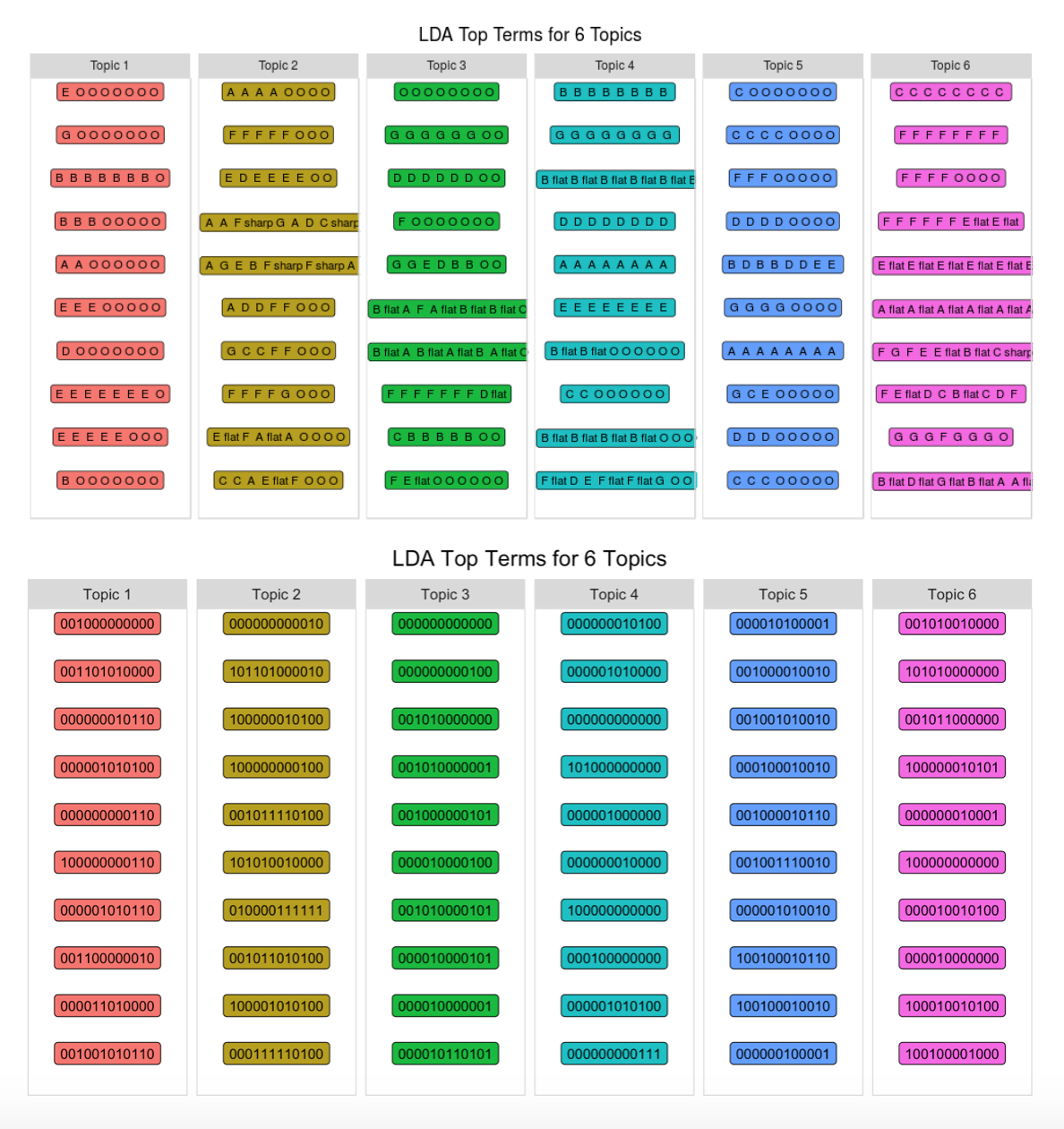}
  \caption{Top 10 Tokens in Selected Topic in Two Scenarios}\label{fig:topwords}
 \end{figure}

Figure \ref{fig:topwords} shows the top 10 tokens in the topics from two scenarios. \\

For Measure-Based Scenario, we can see some topics purely natural keys: \\ 
e.g. Topic 1: $[E,O,O,O,O,O,O,O]$ , Topic 5: $[B,D,B,B,D,D,E,E]$. \\
While some topics are very complicated with many sharps and flats in the notes: \\
e.g. Topic 3: $[B\flat,A,F, A\flat,B\flat,B\flat,O,O ]$, Topic 6: $[F,G,F,E,E\flat,B\flat,C\sharp,D]$. \\ 

For Note-Based Scenario, each token is a 12-dimension vector indicating which of the pitch are "on" in certain measure. Some of the topics contains many active notes: \\ 
e.g. In Topic 2, some tokens have at most 7 active pitches.  \\ 
While some topics are very silent with only few active notes: \\ 
e.g. In Topic 4 most pitches are mute, tokens have at most 3 active pitches. 

\begin{figure}[H]
 \centering
  \includegraphics[scale=0.3]{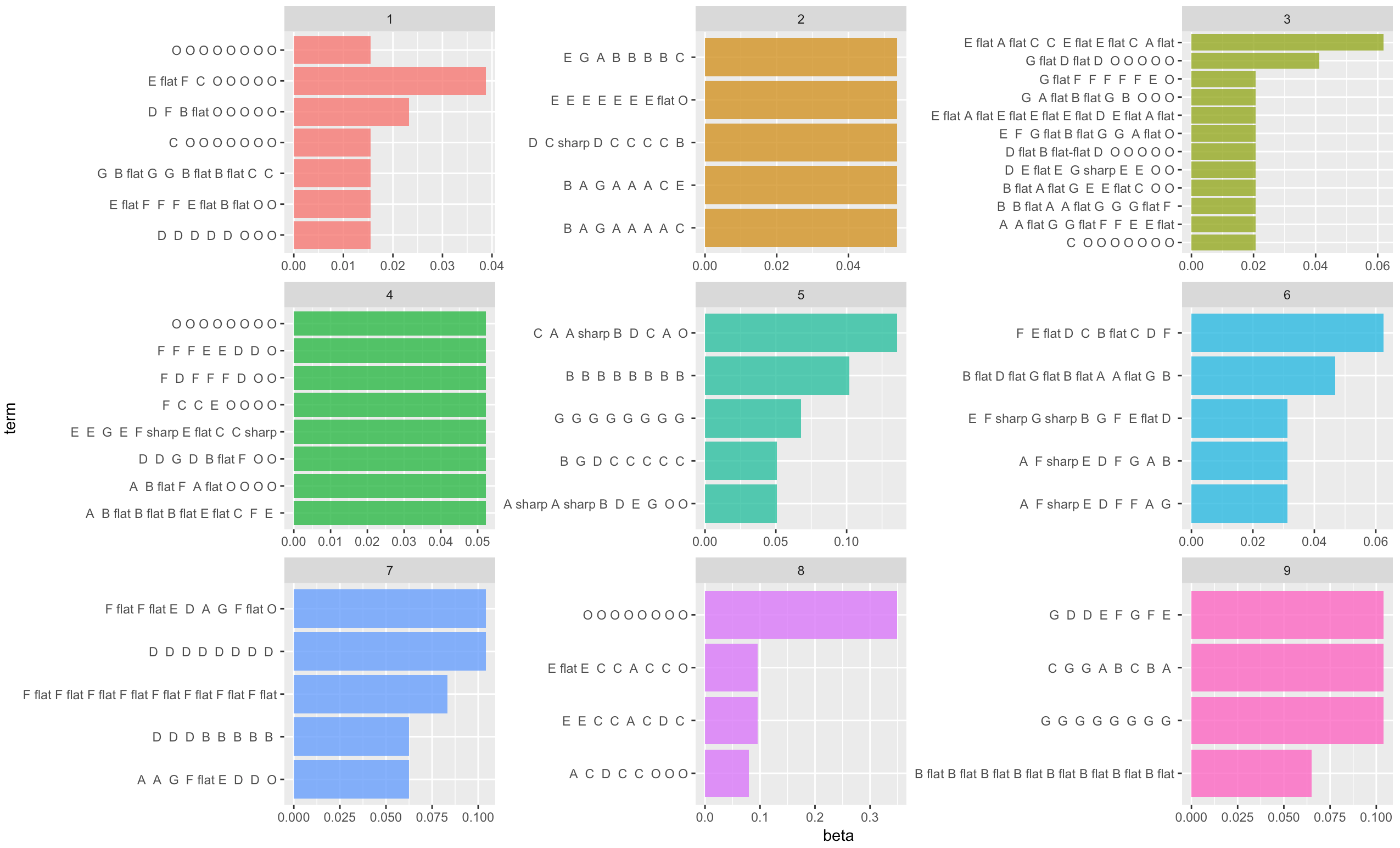}
  \caption{Topic Terms Distribution from Measure-Based Scenario}\label{fig:beta1}
 \end{figure}

Figure \ref{fig:beta1} shows the per-topic per-word probability of Measure-Based Scenario. We can see some topics appear very complicated with most of terms with flat or sharp notes (Topic 3, Topic 4). Some topics are very simple (Topic 8). Some topics contain too many terms with the same probability (Topic 2, Topic 4).

\begin{figure}[H]
 \centering
  \includegraphics[scale=0.2]{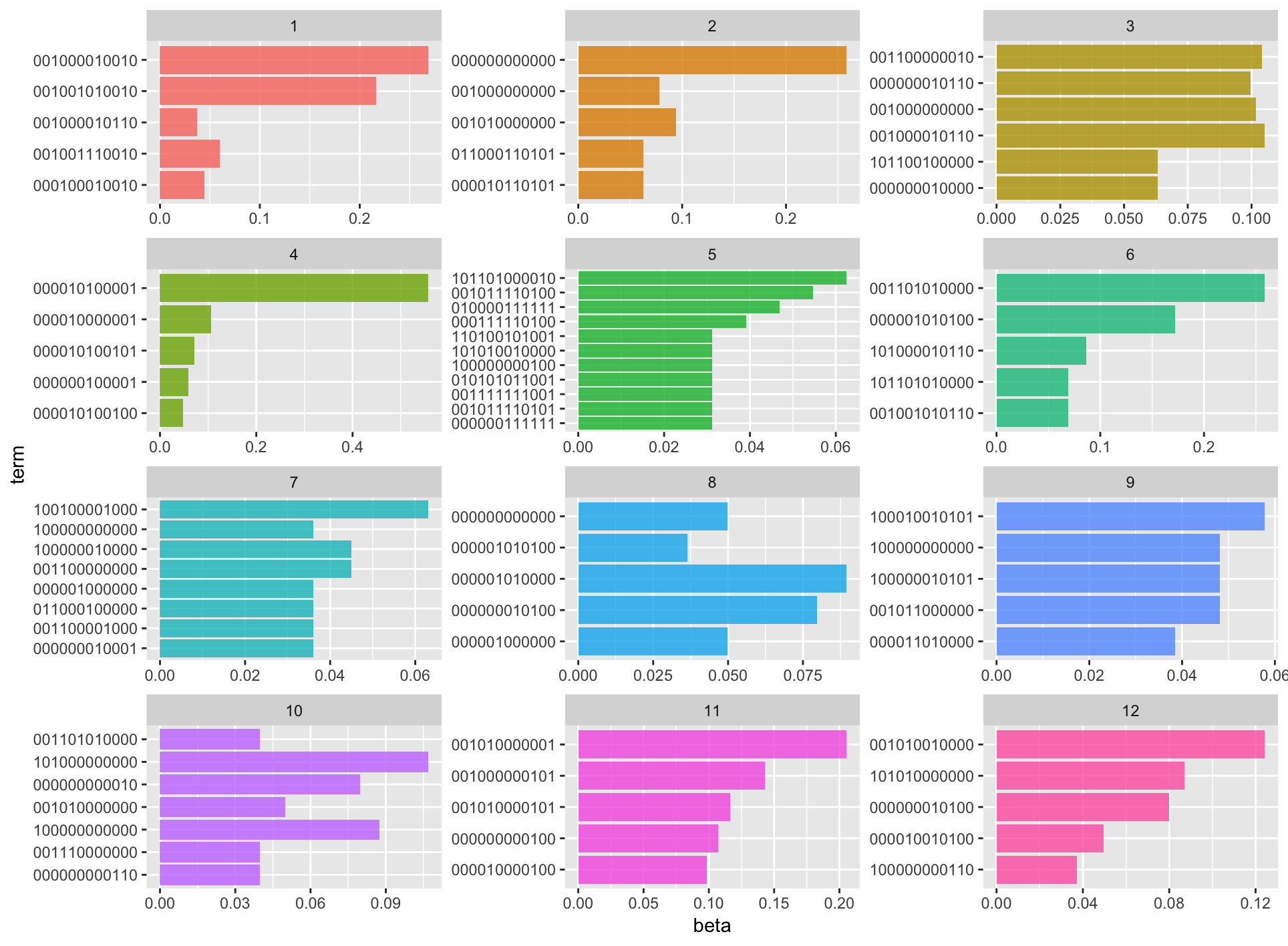}
  \caption{Topic Terms Distribution from Note-Based Scenario}\label{fig:beta2}
 \end{figure}

Figure \ref{fig:beta2} shows the per-topic per-word probability of Note-Based Scenario. Topic 4 and Topic 2 have certain distinctive terms while terms in Topic 9 have fairly similar probability. Further investigation involved musician is needed to better interpret the result. 

\bigskip

\begin{figure}[H]
 \centering
  \includegraphics[scale=0.4]{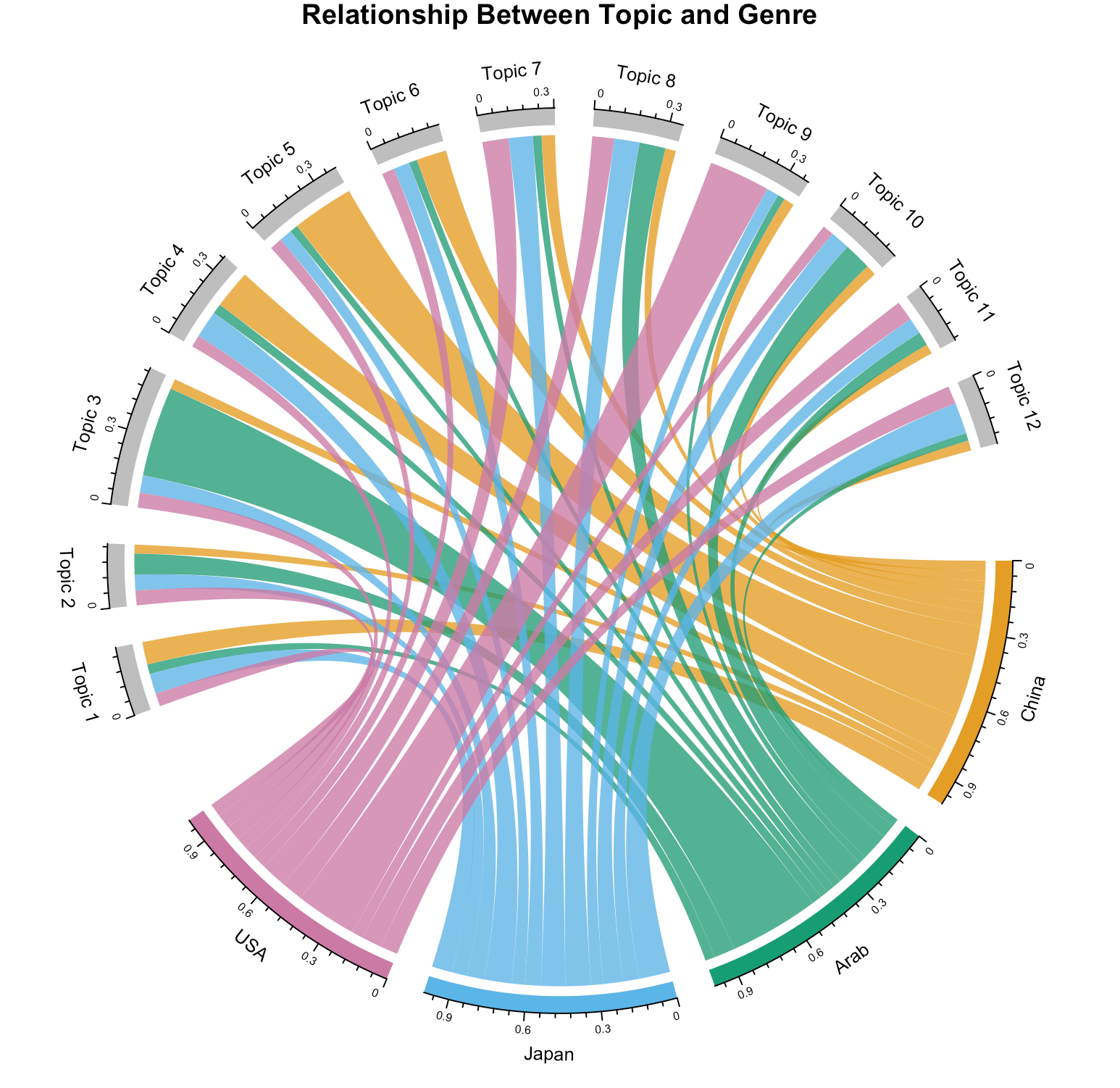}
  \caption{Chord Diagram for Music Genres}\label{fig:mus1}
 \end{figure}

Lastly I draw chord diagram to see some potential relationship between topics learned from topic models and the targeted subjects. \\ 
In Figure \ref{fig:mus1}, we can see:
\begin{itemize}
\item  American songs (Jazz music in this case) are particularly dominant in Topic 9, which has most probable term $[1,0,0,0,1,0,0,1,0,1,0,1]$. It can also be interpreted as pitch class set: $\{C, E, G, A, B\}$, \includegraphics[scale=0.2]{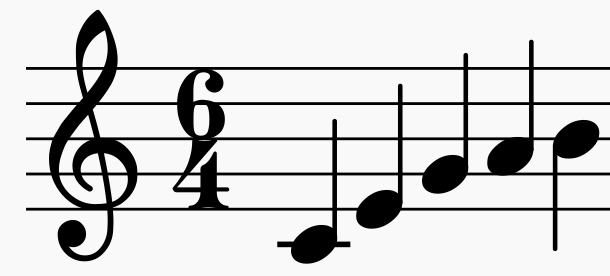}
\item Arabic songs contribute mostly to Topic 3, which has various terms equally distributed (see Figure \ref{fig:beta2}). 
\item Most of Chinese songs attributes to Topic 4 and Topic 5 which contain most probable G major or E minor scale $\{ E, F\sharp, B\}$
\item Japanese songs seem to have similar contribution to every topic. 
\end{itemize}

\begin{figure}[H]
 \centering
  \includegraphics[scale=0.43]{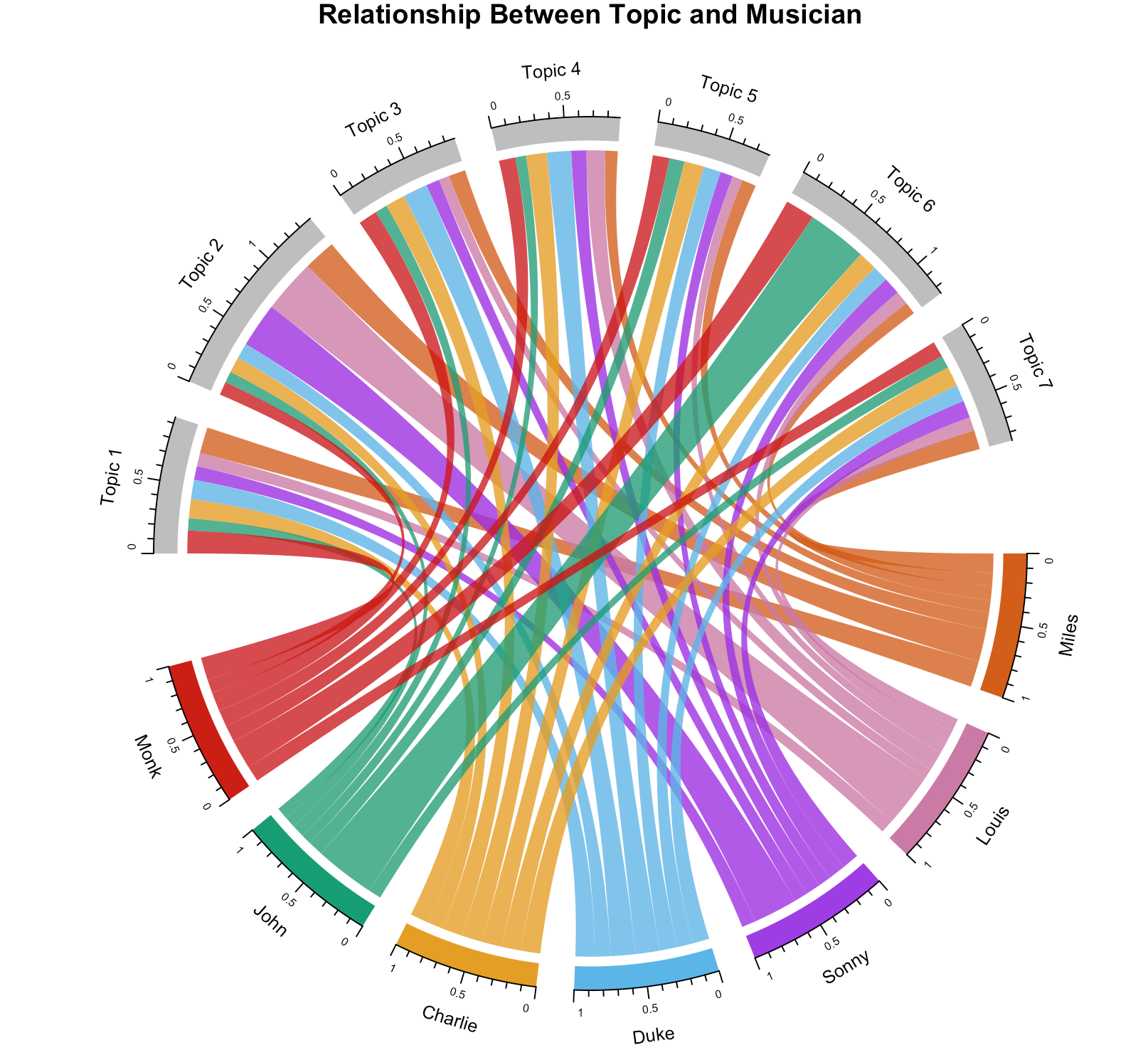}
  \caption{Chord Diagram for Jazz Music}\label{fig:mus2}
 \end{figure}

In Figure \ref{fig:mus2}, we can see:
\begin{itemize}
\item Musician John Coltrane, Sonny Rollins and Louis Armstrong has some certain preference towards certain topics. 
\item Other musicians do not show clear bias to a specific topic. 
\end{itemize}

\section{Conclusion}

\subsection{Summary}

In this paper we create two different representations for symbolic music and transform the music notes from music sheet into matrices for statistical analysis and data mining. Specifically, each song can be regarded as a text body consisting of different musical words. One way to represent these musical words is to segment the song into several parts based on the duration of each measure. Then the words in each song turn out to be a series of notes in one measure. Another way to represent music words is to restructure the notes in each segment based on the fixed 12-dimension pitch class. Both representations have been employed in pattern recognition and topic modeling techniques respectively, to detect music genres based on the collected songs, and figure out the potential connections between musicians and latent topics. \\ 

The predictive performance in pattern recognition for note-based representation turns out to be very good with 88\% accuracy rate in the optimal scenario. We explored several aspects among music genres and musicians to see the hidden associations between different elements. Some genres contain very strong characteristics which make them very easy to detect. Jazz musicians John Coltrane, Sonny Rollins and Louis Armstrong show their particular preference towards certain topics. All these features are employed in the model to help better understand the world of music. 

\subsection{Future Work}
Music mining is a giant research field, and what we've done is merely a tip of the iceberg. Look back to the initial motivation that triggers us to embark on this research work: \textit{Why does music from diverse culture have so powerful inherent capacity to bring people so many different feelings and emotions?} To ultimately find out how to replace human intelligence with statistical algorithms for melody interpretation is still remain to be discovered. 

Several potential studies we would love to continue exploring in the foreseeable future:
\begin{itemize}
\item Facilitate audio music and symbolic music transformation via machine learning technique.
\item Deepen the understanding of musical lexicon and grammatical structure and create the dictionary in a mathematical way. 
\item How to derive representations for smooth recognition of Jazz by statistical learning methods?
\item Apart from notes, can we embed other inherent musical structure such as cadence, tempo to better interpret the musical words?
\item Explore the improvisation key learning (how many keys do the giants of jazz tended to play in, and what are those keys).
\item Musical harmonies and its connection with elements of mood.
\end{itemize}

\section*{Acknowledgments}
We would like to show our gratitude to Dr. Jonathan Kruger, Dr. Evans Gouno, Mrs. Rebecca Ann Finnangan Kemp, Dr. David Guidice for sharing their pearls of wisdom with us during the personal communication on music lexicon.

Special big thank goes to musicians: Lizhu Lu from Eastman School of Music, Gankun Zhang from Brandon University School of Music, Dr. Carl Atkins from Department of Performance Arts \& Visual Culture, and Professor Kwaku Kwaakye Obeng from Brown University, for their encouragement and technical supports in music thoery all the time.

Qiuyi Wu thanks RIT Research \& Creativity Reimbursement Program for partially sponsoring this work to have it possibly presented in Joint Statistical Meetings (JSM) this year in Vancouver. She appreciates supports from International Conference on Advances in Interdisciplinary Statistics and Combinatorics (AISC) for NC Young Researcher Award this year. She thanks 7th Annual Conference of the Upstate New York Chapters of The American Statistical Association (UP-STAT) for recognizing this work and offering her Gold Medal for Best Student Research Award this year.

%This is where your bibliography is generated. Make sure that your .bib file is actually called library.bib
\bibliography{library}

%This defines the bibliographies style. Search online for a list of available styles.
%\bibliographystyle{abbrv}

\nocite{*}
%\printbibliography[heading=bibintoc]

\end{document}